\documentclass[twocolumn]{svjour3}

\RequirePackage{fix-cm}
\usepackage{mathptmx}
\journalname{Stat Comput}

\usepackage{amsmath, amssymb}
\usepackage{verbatim}
\usepackage{caption}
\usepackage{graphicx}
\usepackage{placeins}
\usepackage{multirow}
\usepackage[numbers]{natbib}
\usepackage[noblocks]{authblk}
\usepackage[usenames,dvipsnames]{color}
\newcommand{\ech}{\color{black}  ~}    

\newcommand{\bfy}{{\bf y}}
\newcommand{\bea}{\begin{eqnarray*}}
\newcommand{\eea}{\end{eqnarray*}}

\DeclareMathOperator{\argmin}{argmin}

\begin{document}

\title{Laplace Based Approximate Posterior Inference for Differential Equation Models}

\author{Sarat C. Dass         \and
        Jaeyong Lee \and
        Kyoungjae Lee \and
        Jonghun Park
}
\institute{Sarat C. Dass \at
              Department of Fundamental and Applied Science, Universiti Teknologi PETRONAS
           \and
           Jaeyong Lee \at
          Department of Statistics, Seoul National University
          \and
          Kyoungjae Lee \at
          Department of Statistics, Seoul National University \\
          1 Gwanak-ro, Gwanak-gu, Seoul, 151-747, Korea \\
          Tel.: +82-2-880-8138\\
          \email{leekjstat@gmail.com}
          \and
          Jonghun Park \at
          Department of Industrial Engineering, Seoul National University
}

\date{Received: 10 July 2015 / Accepted: date}

\maketitle

\begin{abstract}
Ordinary differential equations are arguably the most popular and useful mathematical tool for describing physical and biological processes in the real world. Often, these physical and biological processes are observed with errors, in which case the most natural way to model such data is via regression where the mean function is defined by an ordinary differential equation believed to provide an understanding of the underlying process. These regression based dynamical models are called differential equation models. Parameter inference from differential equation models poses computational challenges mainly due to the fact that analytic solutions to most differential equations are not available. In this paper, we propose an approximation method for obtaining the posterior distribution of parameters in differential equation models. The approximation is done in two steps. In the first step, the solution of a differential equation is approximated by the general one-step method which is a class of numerical numerical methods for ordinary differential equations including the Euler and the Runge-Kutta procedures; in the second step, nuisance parameters are marginalized using Laplace approximation. The proposed Laplace approximated posterior gives a computationally fast alternative to the full Bayesian computational scheme (such as Makov Chain Monte Carlo) and produces more accurate and stable estimators than the popular smoothing methods (called collocation methods) based on frequentist procedures. For a theoretical support of the proposed method, we prove that the Laplace approximated posterior converges to the actual posterior under certain conditions and analyze the relation between the order of numerical error and its Laplace approximation. The proposed method is tested on simulated data sets and compared with the other existing methods.
\keywords{Ordinary differential equation \and  posterior computation \and Laplace approximation}
\end{abstract}

\section{Introduction}\label{sec:1}
Ordinary differential equations (ODEs) are arguably the most commonly used mathematical tool for describing physical and biological processes in the real world. Popular examples include Lotka-Volterra equation (Alligood et al. 1997)\nocite{Alligood97}, SIR (Susceptible, Infected, Recovered) model (Kermack and McKendrick 1927)\nocite{Kermack27}  and the continuously stirred tank reactor (CSTR) model (Schmidt 2005)\nocite{Schmidt05}. The Lotka-Volterra equation is the differential equation describing the dynamics of predator-prey systems. The SIR model is an ODE model for disease epidemic describing the relation among the numbers of susceptible, infected and recovered individuals in a closed population. The CSTR model describes the surface temperature changes of an object at a rate proportional to its relative temperature to the surroundings. These are just a few examples of ODEs.

The ODE model is the nonlinear regression model whose regression function is expressed as the solution of an ODE. The ODE model depicts the statistical situation of most applications where the parameters of an ODE need to be estimated based on the noisy data.  The statistical inference for ODE model, however, poses computational challenges mainly due to the lack of analytical solutions for most ODEs.

Bard (1974)\nocite{BardYonathan74} suggested to minimize an objective function, a suitable measure of lack of fit, in which the solution of ODE is approximated by numerical integration. The minimization is carried out by a gradient-based method. But the solutions are often divergent, stay at a local minimizer and  are sensitive to initial values (Cao et al. 2011).\nocite{cao11}

Varah (1982)\nocite{Varah82} proposed an estimation method with the following two steps: in the first step, the regression function is expressed by cubic splines with fixed knots and estimated by least squares method using the data; in the second step, the parameters of the ODE are estimated by minimizing a distance measure between the ODE and the estimated regression function in the first step. Ramsay and Silverman (2005)\nocite{RamsaySilverman05} introduced a two step iteration method where the first step of Varah is modified to a penalized least squares method in which a roughness penalty term is introduced to measure the difference between the ODE and the estimated mean function.

The parameter cascading method was proposed in Ramsay et al. (2007)\nocite{RamsayHooker07}. In the parameter cascading approach, parameters are grouped into (1) regularization parameters, (2) parameters in ODE and (3) regression coefficients in the basis expansion of the regression function. Parameters in each of the three groups are estimated in sequence. First, the regression coefficients are estimated given the structural parameters and regularization parameters, then the structural parameters are estimated given the regularization parameters, and finally the regularization parameters are estimated based on minimizing penalized least squares.

Gelman et al. (1996)\nocite{Gelman96} proposed a Bayesian computational method for the inference of pharmacokinetic models. Huang et al. (2006)\nocite{HuangLiu06} suggested a hierarchical Bayesian procedure for the estimation of parameters in a longitudinal HIV dynamic system. As it turns out, Bayesian computational schemes for ODE models using Markov Chain Monte Carlo type procedures as in these two papers result in even bigger challenges. Each time the parameters are sampled from a candidate distribution, numerical integration of ODE needs to be invoked to evaluate the full likelihood. Campbell (2007)\nocite{Campbell07} adopted the collocation method to obtain an approximation to the regression function expressed by a differential equation as in Ramsey et al. (2007). The collocation method was subsequently combined with parallel tempering (Geyer 1992)\nocite{Geyer92} to overcome instability of the posterior surface. Incorporating tempering overcomes instabilities but slows down computational speed significantly.  Recently, Gaussian processes (GP) have been used to avoid the heavy computation of the numerical integration. Dondelinger et al. (2013) introduced the adaptive gradient matching (AGM) approach which has a link to numerical integration but without the corresponding high computational cost.
Wang and Barber (2014) introduced the Gaussian process-ODE (GP-ODE) approach which provides a generative model and simpler graph model than AGM approach. Actually, GP-ODE approach makes an approximation to allow a generative and proper graph model as Macdonald et al. (2015) pointed out.

In this paper, to speed up the Bayesian computations, we propose a Laplace approximated procedure (LAP) for posterior inference in differential equation models. The marginal posterior density of the ODE parameter is computed by the Laplace approximation (LA), in which the regression function is approximated by a one-step numerical solver of ordinary differential equations. We use the Euler and the fourth order Runge-Kutta procedures for illustrations. Finally, posterior inference is carried out by grid sampling or griddy Gibbs sampling from the marginal posterior of the ODE parameters depending on its dimension.

The proposed method has the following advantages. First, for an ODE model with the parameter dimension less than or equal to four, the posterior computations utilizes the Monte Carlo method (not the Markov Chain Monte Carlo method) based on independent sampling; thus, its posterior sampling is significantly faster than methods utilizing full Bayesian computations. Even for moderate parameter dimensions, the LAP runs and produces results within an acceptable computational time frame.

The second advantage is that the LAP produces more accurate parameter estimates compared to the other existing methods.  In a simulation study, we compared the LAP with the parameter cascading method (Ramsay et al., 2007), the delayed rejection adaptive Metropolis algorithm (Haario et al., 2006), GP-ODE approach (Wang and Barber, 2014) and AGM approach (Dondelinger et al., 2013)\nocite{dondelinger2013ode}. In the FitzHugh-Nagumo model where the regression function changes more rapidly, the LAP estimator has better performance than the delayed rejection adaptive Metropolis (DRAM), GP-ODE and AGM approach in the sense of the root mean squared error (rmse) and the log-likelihood at the parameter estimates. The performance of LAP is comparable to the parameter cascading (PC) method in the same sense. The latter criteria judges whether the chosen procedure achieves a parameter estimate that is close to the maximum likelihood by ascertaining the corresponding log-likelihood value.

Third, inference based on the LAP is numerically stable. Frequentist methods need to maximize the log-likelihood surface which has many ripples. So, depending on the starting points, optimization algorithms can be trapped in local maximums. However, in many examples, the ripples of the log-likelihood surface occur at the periphery of the parameter space and disappear from the likelihood surface when the sample size $n$ becomes large.

The rest of the paper is organized as follows. In Sect. \ref{sec:2}, we lay out inference framework of the differential equation models and the priors considered in this paper. The proposed posterior computations are described in Sect. \ref{sec:3}. In Sect. \ref{sec:4}, we prove  that the approximated posterior converges to the true posterior under certain regularity conditions. In Sect. \ref{sec:5}, using the simulated data sets from three models, we examine the quality of the LA based posterior. In the examples we considered, inference based on LAP generates stable and accurate approximations of the true posterior. We apply the LAP to a real data set, U.S. Census data in Sect. \ref{sec:6}. Discussions are presented in Sect. \ref{section:discussion} whereas details of computations and technical results are relegated to the Appendix.

\section{Regression model defined by ODE}\label{sec:2}
We consider the regression model
\begin{eqnarray*}
y(t) &=& x(t) + \epsilon(t),
\end{eqnarray*}
where $y(t)$ is a $p$-dimensional vector of observation at time $t \in [T_0,T_1], 0 \le T_0 < T_1 < \infty$ and $\epsilon(t)$ represents an error term assumed to arise from $N_p(0, \sigma^2 I_p)$ with $\sigma^2 > 0$ where $N_p(\mu, \Sigma)$ denotes the $p$-dimensional normal distribution with mean $\mu$ and covariance matrix $\Sigma$. The regression function, $x(t)$, of the regression model is defined as the solution of a differential equation
\begin{equation}\label{ode}
\dot{x}(t) = f(x, u, t ; \theta), \,\, t \in [T_0, T_1],
\end{equation}
where $f$ is a $p$-dimensional smooth function of $x(t)$, known input function $u(t)$, time $t$, and the unknown parameter $\theta \in \Theta \subseteq R^{q}$ with $q \ge 1$; $\dot{x}(t)$ denotes the first derivative of $x(t)$ with respect to time $t$. The function $x$ is determined by the initial value of $x$, $x(T_0)$, $\theta$ and the function $u(\cdot)$. The unknown parameter $\theta$ needs to be estimated from observed data on $y(t)$s and $u(t)$s which are given at certain pre-specified time points.

We assume that observed data is collected at the time points $T_0 \leq t_1 < t_2 < \ldots < t_n \leq T_1$. Letting $y_i = y(t_i)$, $x_i = x(t_i)$ and $\epsilon_i = \epsilon(t_i)$, we have the following regression model
\begin{equation}\label{model}
y_i = x_i + \epsilon_i, \,\, i=1,2,\ldots, n,
\end{equation}
where $\epsilon_i$ are drawn independently from $N_p(0, \sigma^2 I_p)$.

The value of each $x_i$, $i=1,2,\cdots,n$, is determined by the initial value $x_1$,  $\theta$ and $u(\cdot)$ based on the differential equation model (\ref{ode}). When we need to emphasize this dependence,  we  will denote $x_i$ by $x_i \equiv x_i( \theta, x_1, u)$ or $x_i(\theta, x_1)$ if $x$ is not dependent on $u$. For simplicity of exposition, the input function $u(t)$ is not considered further in the rest of the paper, but analysis based on a known input function can be easily accommodated into our inference framework. 

The differential equation (\ref{ode}) involves only the first order derivatives, but can be used to describe those with higher order derivatives. For example, consider a second order equation $\ddot{x}(t) = f(\dot{x}, x ,  t ; \theta)$. By introducing $z(t) = \dot{x}(t)$, the differential equation model can be expressed as
\begin{eqnarray*}
\dot{X}(t) \equiv \binom{\dot{x}(t)}{\dot{z}(t)} = \binom{f(x,t; \theta)}{f(z,x,t ; \theta)} \equiv F(X,t ; \theta)
\end{eqnarray*}
where $X(t) \equiv (x(t)^T, z(t)^T)^T$ is now a vector with an added component for the dynamics of $z(t)$. Since any higher order differential equation models can be converted into a first order differential equation model based on adding extra dynamical systems and variables, without loss of generality, we consider only the first order differential equation models for developing our inference procedures in the remainder of this paper.

In the model (\ref{ode}) and (\ref{model}), there are three unknowns, $x_1$, $\theta$ and $\sigma^2$, whose priors are denoted by $\pi(x_1 \mid \sigma^2)$, $\pi(\theta)$ and $\pi(\sigma^2)$ (or $\pi(\tau^2)$ with $\tau^2 = 1/\sigma^2$), respectively.
In the following, we will take the following specific priors for $\tau^2$ and $x_1$:
\begin{eqnarray}
\label{priortau2} \tau^2 & \sim & Gamma(a,b) \\
\label{priorx1} x_1 \mid \tau^2 & \sim &  N_p(\mu_{x_1}, c \tau^{-2} I_p),
\end{eqnarray}
where $ c > 0$ and $Gamma(a, b)$ is the gamma distribution with parameters $a, b >0$ and mean $a/b$.  The prior selection for $(\tau^{2},x_1)$ is guided by conjugacy considerations which enable components of the posterior to be integrated in closed form. One may select other types of priors for $(\tau^{2},x_1)$. However, for large sample sizes, like the ones considered in this paper, the impact of these priors will be minimal since most of the inference will be driven and guided by the likelihood component of the posterior.

\section{Posterior Computation}\label{sec:3}
\subsection{Posterior of $\theta$, $\tau^2$ and $x_1$}\label{sec:3.1}
The full joint posterior of  $\theta$, $x_1$ and $\tau^2$ given the observations ${\bf y}_n = (y_1, y_2, \ldots, y_n)^T$ has the expression $\pi(\theta, \tau^2, x_1 \mid {\bf y}_n)$
\begin{eqnarray*}
\quad &\propto& p({\bf y}_n \mid \theta, \tau^2, x_1) \pi(x_1 \mid \tau^2)\pi(\tau^2) \pi(\theta)  \\
&=& \left[\prod_{i=1}^n det(\tau^{-2} 2\pi I_p)^{-1/2} e^{-\frac{\tau^2}{2} \|y_i - x_i(\theta, x_1) \|^2}\right] \\
&& \quad\times \,det(2\pi c \tau^{-2} I_p )^{-1/2} e^{-\frac{\tau^2}{2c} \|x_1 - \mu_{x_1}\|^2} \\
&&\quad\times \,\frac{b^a}{\Gamma(a)} (\tau^{2})^{a-1} e^{-b\tau^2} \times \pi(\theta) \\
& \propto & (\tau^2)^{\frac{1}{2} (n+1)p +a -1} \times  e^{-\frac{\tau^2}{2} (ng_n(x_1,\theta)+ \frac{\|x_1 - \mu_{x_1}\|^2}{c} + 2b) } \pi(\theta), \nonumber
\end{eqnarray*}
where
$g_n(x_1) = g_n(x_1,\theta) = \sum_{i=1}^n \|y_i - x_i(\theta,x_1)\|^2 /n$ and $\|x\|$ denotes the Euclidean norm of the vector $x$, and $\pi(\theta)$ is {\em any} prior on $\theta$. The choice of $\pi(\theta)$ can be arbitrary as it does not affect the inference on $\theta$ for large sample sizes $n$ as is well known. 

In most cases, $\theta$ and $\tau^2$ are the parameters of primary interest whereas $x_1$ is the nuisance parameter. The details of obtaining the posterior distributions of $\theta$ and $\tau^{2}$ are outlined as follows: In the first step, the posterior of $\theta$ and $\tau^2$, $\pi(\theta, \tau^2 \mid {\bf y}_n)$, is obtained by marginalizing (i.e., integrating out) $x_1$. In this marginalization step, two approximations are implemented:  (i) a one-step numerical method for calculating each $x_i$, $i=1,2,\cdots,n$, and (ii) the Laplace method for integrating out $x_1$. In the next step, as a consequence of conjugacy, it can be shown that the posterior of $\tau^2$ given $\theta$ and ${\bf y}_n$ follows a gamma distribution, which has the advantage that it can be easily and directly sampled from. In the third step, after marginalizing $\tau^{2}$, $\theta$ is sampled from its posterior distribution, $\pi(\theta \mid {\bf y}_n)$, using either grid sampling or griddy Gibbs sampling depending on its dimension $q$. By eliminating $x_1$ and $\tau^{2}$ from the full posterior in the first two stages above, we reduce the dimension of the posterior from $p+q+1$ to $q$, making it easier for thorough exploration of its surface using grid based sampling as in the third stage.

\subsection{Marginalization of $x_1$: Joint posterior of $\theta$ and $\tau^{2}$}\label{sec:3.2}

In the marginalization of $x_1$, we use two approximations. In the first approximation, $x_i(\theta, x_1)$ is successively approximated by a numerical procedure:
$$x_i \approx x_{i-1} + (t_i - t_{i-1})\phi( x_{i-1}, t_{i-1} ; \theta),~~ i=2,\ldots, n,$$
where different forms of $\phi$ represent different numerical solvers of differential equation.
For example, the Euler method is represented by
$$\phi( x_{i-1}, t_{i-1} ; \theta) = f(x_{i-1}, t_{i-1} ; \theta) ;$$
while the 4-th order Runge-Kutta is represented by
\begin{equation}\label{RK4}
\phi( x_{i-1}, t_{i-1} ; \theta) = {1 \over 6}(k_{i-1,1} + 2k_{i-1,2} + 2k_{i-1,3} + k_{i-1,4}),
\end{equation}
where
\bea
k_{i-1,1} &=& f(x_{i-1}, t_{i-1} ; \theta) , \\
k_{i-1,2} &=& f(x_{i-1} + {1 \over 2}k_{i-1,1} , t_{i-1} + {1\over2}(t_i - t_{i-1}) ; \theta), \\
k_{i-1,3} &=& f(x_{i-1} + {1\over2}k_{i-1,2},t_{i-1} + {1\over2}(t_i - t_{i-1}) ; \theta), \\
k_{i-1,4} &=& f(x_{i-1} + k_{i-1,3} , t_{i} ; \theta).
\eea
Let $h = \max_{2 \le i \le n} (t_i - t_{i-1})$ and $x^h$ be the approximation of $x$.
The global error of the numerical method is defined by
$$\sup_{t\in [T_0,T_1]} \| x(t) - x^h(t)\| .$$
If the global error is $O(h^K)$ for some integer $K$, we call $K$ to be the order of the numerical method. Under some smoothness conditions, the order of the 4th order Runge-Kutta numerical procedure (given in \eqref{RK4}) is $K=4$ (Mathews and Fink 2004; S\"uli 2014).\nocite{mathews2004numerical} \nocite{suli14}

In the second approximation, we integrate out $x_1$ based on its prior $\pi(\,x_1\,|\,\tau^{2})$ defined in (\ref{priorx1}) and full likelihood using Laplace approximation for the corresponding integral. Using results from Tierney and Kadane (1986) and Azevedo-Filho and Shachter (1994),\nocite{AzevedoRoss94} \nocite{Tierney86} the marginal likelihood of $\theta$ and $\tau^2$ can be approximated by
\begin{eqnarray*}\label{x1out}
L(\theta, \tau^2) &=& \int \pi(x_1 \mid \tau^2) L(\theta, \tau^2, x_1) dx_1 \nonumber \\
&\propto& \int (\tau^2)^{(n+1)p/2} e^{- \frac{\tau^2}{2} \left(\,n g_n(x_1) + \frac{\| x_1 - \mu_{x_1} \|^2}{c}\,\right)} dx_1 \\
&\propto&  (\tau^2)^{(n+1)p/2}\,  e^{-\frac{\tau^2}{2} \,u(\theta) }\,\, det\,\left(\,n \ddot{g}_n(\hat{x}_1) + \frac{2}{c} I_p \,\right)^{-1/2}\\
&& \times \,(\tau^2)^{-p/2}\left(1+O(n^{-3/2})\right)   \\
&=& (\tau^2)^{np/2}\,  e^{-\frac{\tau^2}{2} \,u(\theta) -\frac{1}{2}\, v(\theta) } (1+O(n^{-3/2})),
\end{eqnarray*}
where
\begin{eqnarray*}
\ddot{g}_n(x_1) &=& \frac{\partial^2\,g_n(x_1,\theta)}{\partial x_1^2},\,\,\hat{x}_1 \equiv \hat{x}_1(\theta)\,\mbox{is given by}   \\
\hat{x}_1(\theta) &=& \underset{x_1}{\argmin} \big( ng_n(x_1, \theta) + \frac{\| x_1 - \mu_{x_1} \|^2}{c} \big), \\
u(\theta) & =& n g_n(\hat{x}_1) + \frac{\| \hat{x}_1 - \mu_{x_1}\|^2}{c}, \mbox{ and}  \\
v(\theta) & =&   \log\, det\left(  n \ddot{g}_n(\hat{x}_1) + \frac{2}{c}I_p\right).
\end{eqnarray*}

It follows from the last expression for $L(\theta,\tau^{2})$ that the approximate posterior of $\theta$ and $\tau^2$ given ${\bf y}_n$, based on independent priors $\pi(\theta)$ and  $Gamma(a,b)$ on $\theta$ and $\tau^{2}$, respectively, is given by
\begin{eqnarray}\label{posttt}
\pi(\theta, \tau^2 \mid {\bf y}_n) &\propto& \pi(\theta) \times (\tau^2)^{\frac{np}{2} +a - 1}  e^{-\tau^2 (\frac{1}{2}\, u(\theta)+b)}  \\
&&\times \,\, det\left( \, n \ddot{g}_n(\hat{x}_1) + \frac{2}{c}I_p\,\right)^{-1/2}\nonumber.
\end{eqnarray}

Details for the computation of $\ddot{g}_n(x_1)$ is given in the Appendix.  We used the gradient descent and Newton-Raphson procedures for obtaining the maximizer $\hat{x}_1(\theta)$ in our examples.

\subsection{Marginalization of $\tau^{2}$: Posterior of $\theta$}\label{sec:3.3}

We note from equation (\ref{posttt}) that the posterior of $\tau^2$ given $\theta$ and ${\bf y}_{n}$ is proportional to
$$
\pi(\tau^2 \mid \theta , {\bf y}_n) \propto  (\tau^2)^{\frac{np}{2} + a - 1}  e^{-\tau^2 \left(\frac{1}{2}\, u(\theta)+b\,\right)}; 
$$
thus, the conditional posterior distribution of $\tau^{2}$ given $\theta$ and ${\bf y}_{n}$ is given by 
$$\tau^2 \mid \theta, {\bf y}_n \sim Gamma(a^*, b^*),$$ where $a^*  =  np/2 +a$ and $b^*  =  u(\theta)/2 +b$. Now, by integrating out $\tau^2$ from the product of $L(\theta, \tau^2)$ and the prior of $\tau^2$, we get the marginal likelihood of $\theta$ given by
\begin{eqnarray*}
L(\theta) & \propto & \int (\tau^2)^{\frac{np}{2} +a - 1} \, e^{-\tau^2 \left(\frac{1}{2} u(\theta)+b\right)} d\tau^2 \\ && \times \,\,  det\,\left(  n \ddot{g}_n(\hat{x}_1) + \frac{2}{c}I_p\,\right)^{-1/2} \\
& = & \frac{\Gamma(\frac{np}{2} +a )}{\left(\frac{1}{2} u(\theta)+b\right)^{\frac{np}{2} +a }}  \,det\,\left(n \ddot{g}_n(\hat{x}_1) + \frac{2}{c}I_p\,\right)^{-1/2}.
\end{eqnarray*}
Consequently, the posterior of $\theta$ is
\begin{equation}\label{post-theta}
\pi(\theta\mid {\bf y}_n)  ~\propto~   \frac{\pi(\theta)}{\left(\frac{1}{2}\, u(\theta)+b\right)^{\frac{np}{2} + a } \,det\,\left(\,n \ddot{g}_n(\hat{x}_1) + \frac{2}{c}I_p\,\right)^{1/2}}.
\end{equation}

To numerically approximate $\pi(\theta \mid {\bf y}_n)$, we propose grid sampling or griddy Gibbs sampling depending on the dimension of $\theta$. When the dimension $q$ of $\theta$ is not large (say, $q \le 4$), the grid sampling is conceptually simple and numerically fast. When $q$ is relatively large, we recommend the griddy Gibbs sampling.


\subsection{Posterior sampling of $\theta$}\label{sec:3.4}
When $q \leq 4$, we recommend the grid sampling to sample $\theta$ from the marginal posterior $\pi(\theta \mid \bfy_n)$ of $\theta$ in \eqref{post-theta}.
Let ${\cal {G}}_{\Theta} \subset \Theta$ be a grid set that covers $\Theta$ and let $\pi^d(\theta \mid {\bf y}_n)$ be the discrete distribution with support  ${\cal {G}}_{\Theta}$ whose value at $\theta \in {\cal {G}}_{\Theta}$ is proportional to $\pi(\theta \mid {\bf y}_n)$.
We will sample $\theta$ from $\pi^d(\theta \mid {\bf y}_n)$.

In practice, the choice of the grid matrix ${\cal {G}}_{\Theta}$ can be a nontrivial task (Joshi and Wilson 2011\nocite{joshi11}). To choose a grid set, we adopt the reparametrization technique used by Rue et al. (2009).\nocite{Rue09} Let $\theta^0$ be the initial guess for the center of the grid set, and let $\hat{\Sigma} = H^{-1}$ where $H$ is the negative Hessian matrix of $\pi(\theta | \bfy_n)$ at $\theta^0$.  If $H$ is not a positive definite matrix, we replace the negative eigenvalues of $H$ with the minimum positive eigenvalue of it. We express $\theta$ with a standardized variable $z$ by
$$\theta(z) = \theta^0 + U D^{1/2} z$$
where $\hat{\Sigma}$ is diagonalized with $\hat{\Sigma} = UD U^T$, $U = (u_1,\ldots, u_q)$ and $D = diag( \lambda_j)$. $\lambda_j$ is the eigenvalue of $\hat{\Sigma}$, and $u_j$ is the corresponding eigenvector, $j=1,2, \ldots, q$. The grid points are selected for the parametrization of $z$.
We recommend the two step approach in choosing the range of the grid points.
In the first step, the grid points for the $i$th coordinate $z_i$ is chosen by dividing $[-4, 4]$ into $2M_1$ equal length intervals resulting $2M_1 + 1$ points. Note $[-4, 4]$ comes from the rough normal approximation. For each $(2M_1+1)^q$ grid points, we evaluate $\pi(\theta(z_i) | \bfy_n)$, $i=1,2, \ldots, (2M_1+1)^q$. With these values, we determine the range $[A_i, B_i]$ of each coordinate $z_i$, $i=1,2, \ldots, q$. $A_i$ and $B_i$ are defined by the minimum  and maximum of $z_i$ with $\pi(\theta(z_1, \ldots, z_i, \ldots,  z_q) | \bfy_n)$ $> \eta$ where $\eta$ is a small number close to $0$. In our examples, we used $\eta = 10^{-5}$.
If the interval $[-4, 4]$ is not big enough to contain the mass of the posterior and $A_i$ and $B_i$ can not be selected, we perform the first step one more time with larger interval than $[-4, 4]$. The larger interval can be obtained by approximating the marginal posterior with normal density with larger standard deviation.

After $[A_i, B_i]$ are chosen, we move to the second step and determine the grid points for accurate computation.
The purpose of the first step is to determine the grid set, and  $M_1$ is chosen as a  small positive integer such that $(2M_1 + 1)^q$ is not overwhelmingly large computationally. In our examples, we used $M_1 = 5$.

In the second step, $[A_i, B_i]$ is divided into $2M_2$ intervals of equal length. The discrete approximation of  the posterior is constructed by evaluating the posterior at $(2M_2 +1)^q$ grid points. Grid sampling is done first by sampling $\theta^{(i)}$ from the discrete approximation and the conditionally on $\theta^{(i)}$, $\tau^2$ is sampled from $Gamma(np/2 +a, u(\theta^{(i)})/2 + b)$. In our examples, we used $M_2 = 15$ or $25$.
Note that the samples from this algorithm are independent samples. When $q$ is not very large, the algorithm is very fast.

We summarize the algorithm below.
\begin{enumerate}
\item (Step 1: Reparameterization step) \newline Compute the initial guesses of the center $\theta^0$ and of the posterior covariance $\hat{\Sigma}$. \\ Reparametrize $\theta$ using the standardized variable $z$ by
$$\theta(z) = \theta^0 + U D^{1/2} z$$
where $\hat{\Sigma} = U D U^T$.
\item (Step 2:Finding ranges of $z_i$) \newline For each $z_i$, divide the interval $[-4, 4]$ into $2M_1$ intervals of equal length. Let
\begin{eqnarray*}
A_i & = & \min \{ z_i :  \pi(z_i \mid \bfy_n) \geq \eta \} \\
B_i & = & \max\{ z_i : \pi(z_i \mid \bfy_n) \geq \eta \}.
\end{eqnarray*}
\item (Step 3: Grid sampling)\newline
Divide the intervals $[A_i, B_i]$ into $2M_2$ intervals of equal length and construct grid points.
\begin{enumerate}
\item[1.] For each $\theta \in {\cal {G}}_{\Theta}$, calculate $\pi(\theta |{\bf y}_n)$ using (\ref{post-theta}) and construct $\pi^d(\theta \mid {\bf y}_n)$.
\item[2.] Sample $\theta^{(1)}, \theta^{(2)}, \ldots, \theta^{(N)} \stackrel{iid}{\sim} \pi^d(\theta \mid {\bf y}_n)$.
\item[3.] For each $i=1,2,\ldots, N$, sample \\ ${\tau^2}^{(i)} \sim Gamma(np/2 +a, u(\theta^{(i)})/2 + b)$.
\end{enumerate}
\end{enumerate}

When $q$ is large ($q  \geq 5)$, the construction of the discrete approximation $\pi^d$ by evaluating the posterior at all the grid points can be computationally prohibitive. In this case, we recommend to replace the grid sampling by the griddy Gibbs sampling in the above algorithm.  In the griddy Gibbs sampling, the coordinates of $z$ is sampled from the conditional posterior and it  does not require the evaluation of the posterior at all grid points.

To improve accuracy of the numerical solution of differential equation, we divided the interval $[t_{i-1}, t_{i}]$ to $m$ intervals and added intermittent time points in computing $x$.
If the differential equation is smooth enough, $m = 4$ and $1$ usually suffice for Euler and 4th order Runge-Kutta method not to add  error rate to that of the Laplace approximation, respectively. See Theorem \ref{mrate}.
But in practice sometimes larger values of $m$ are required. We  apply larger values in turn,  and if  the change in the mean of the posterior is less than $0.1\%$, we stopped. In our examples, we used the sequence of $m$ as  $1, 2, 4, 8, 14, 20, 30,\ldots$.

\section{Convergence of the approximated posterior}\label{sec:4}
\subsection{Convergence of the approximated posterior as $m$ increases}\label{sec:4.1}
In this section, we show that the posterior with Laplace approximation and numerical method, $\pi^{LP}_m$, converges pointwise to the true posterior with an relative error of $O(n^{-3/2})$ as $m \to \infty$, under some regular conditions.

For convenience, let $\pi_m \equiv \pi^{LP}_m$.
We assume  $h \equiv t_{i+1} - t_i $ for all $i=2,3,\ldots, n$ and each $[t_{i-1}, t_i]$ is divided into  $m$ segments; thus, the length of one segment is $h/m.$
Let $x^m$ be the approximation of $x$ by numerical method with $m$ segment and $x^m(t_1) = x(t_1)$ for all $m$.

The theorem requires the following assumptions.
\begin{itemize}
\item[A1.] $\{ x(t) : t \in [T_0, T_1] \}$ is a compact subset of $\mathbb{R}^p$;
\item[A2.] $\{ y(t) : t \in [T_0,T_1] \}$ is a bounded subset of $\mathbb{R}^p$;
\item[A3.] the $K$th order derivative of $f(x,t ;\theta)$ with respect to $t$ exists and is continuous in $x$ and $t$, where $K$ is the order of the numerical method $\phi$; and
\item[A4.] the function $ng_n(x_1) + \|x_1 - \mu_{x_1}\|^2/c$ has the unique minimum $\hat{x}_1$.
\end{itemize}
\begin{theorem}\label{Conv}
Suppose that $f(x,t ; \theta)$ is Lipschitz continuous in $x$, and A1 -- A4 hold. Then,  for sufficiently large $n$,
$$\lim_{m\to\infty} \pi_m(\theta,\sigma^2 \mid {\bf y}_n) = \pi(\theta,\sigma^2 \mid {\bf y}_n) \times (1+ O(n^{-3/2})),$$
for all $\theta$ and $\sigma^2$.
\end{theorem}
The proof of theorem is given in Appendix.

\subsection{Suitable rate of step size with respect to sample size}\label{sec:4.2}
In this section, we analyze the relation between the step size $h/m$ and the approximation error rate of the posterior, which is motivated by Xue et al. (2010)\nocite{xue2010sieve}. We assume that the number of the observation goes to infinity and  $h/m = O(n^{-\alpha})$.
The large sample size and small step size give accurate inference, but they may cause heavy computation. We are interested in a reasonable choice of the step size $h/m$ when the sample size $n$ is growing. Here, reasonable choice means that it does not raise the relative error rate $O(n^{-3/2})$ caused by the Laplace approximation.

Let $K$ be the order of the numerical method $\phi$. If we divide intervals $[t_{i-1},t_i]$ into $m$ segments, \\ $\max_{1 \le i \le n} \| x_i - x^m_i \| = O((h/m)^K) = O(n^{-K\alpha})$.

\begin{theorem}\label{mrate}
Suppose that $f(x,t; \theta)$ is Lipschitz continuous in $x$, and $A1-A3$ hold. Let $K$ be the order of the numerical method $\phi$ and $h/m = O(n^{-\alpha})$. If $\alpha \ge 5/(2K)$, then, for sufficiently large $n$,
$$\pi_m(\theta,\tau^2 \mid {\bf y}_n) = \pi(\theta,\tau^2\mid {\bf y}_n) \times (1+ O(n^{-3/2})),$$
for all $\theta$ and $\tau^2$.
\end{theorem}

Theorem \ref{mrate} says that if we set $h/m = O(n^{-5/(2K)})$, the numerical approximation does not raise the order of the relative error caused by the Laplace approximation. Moreover, even if we take $h/m  \ll n^{-5/(2K)}$, it does not reduce the error rate $O(n^{-3/2})$ and only raise the computational cost.


\section{Simulated Data Examples}\label{sec:5}
In this section, we test our LAP inference with data sets simulated from three ODE models. The data are generated with predetermined parameter value $\theta$, the initial value $x_1$ and error variance $\sigma^2$.

For the examples in \ref{NC}, we use both the Euler and the 4th order Runge-Kutta method to approximate ODE solutions.
For the examples in \ref{FN} and  \ref{PP}, we use the 4th order Runge-Kutta method to approximate ODE solutions.
The LAP inference can be extended to other numerical methods by changing the function $\phi(x, t ; \theta)$.

\subsection{Newton's law of Cooling}\label{NC}\label{sec:5.1}
\subsubsection{Model description and data generation}\label{4.1.1}
English physicist Isaac Newton believed that temperature change of an object is proportional to the temperature difference between the object and its surroundings. This intuition is captured by Newton's law of cooling, which is an ODE given by
\begin{eqnarray}\label{NLC}
{\dot x}(t) & = & \theta_1(x(t) - \theta_2),
\end{eqnarray}
where $x(t)$ is the temperature of the object in Celcius at time $t$, $\theta_1$ is a negative proportionality constant and $\theta_2$ is  the temperature of the environment.
See Incropera (2006) for the details. \nocite{Incropera06} The solution of the ODE (\ref{NLC}) is known and is
$$x(t) = \theta_2 - (\theta_2 - x_1)e^{\theta_1 t}$$
where $x_1 \equiv x(0)$. Since the analytic form of the solution is known,
 it is not necessary to resort to the proposed approximate posterior computation method to fit the ODE model with (\ref{NLC}).
We have chosen this example as a testbed for the proposed method. We compare the true posterior without approximation with the approximate posterior obtained by the proposed method.

\begin{figure*}
\centering
		\includegraphics[width=0.65\textwidth]{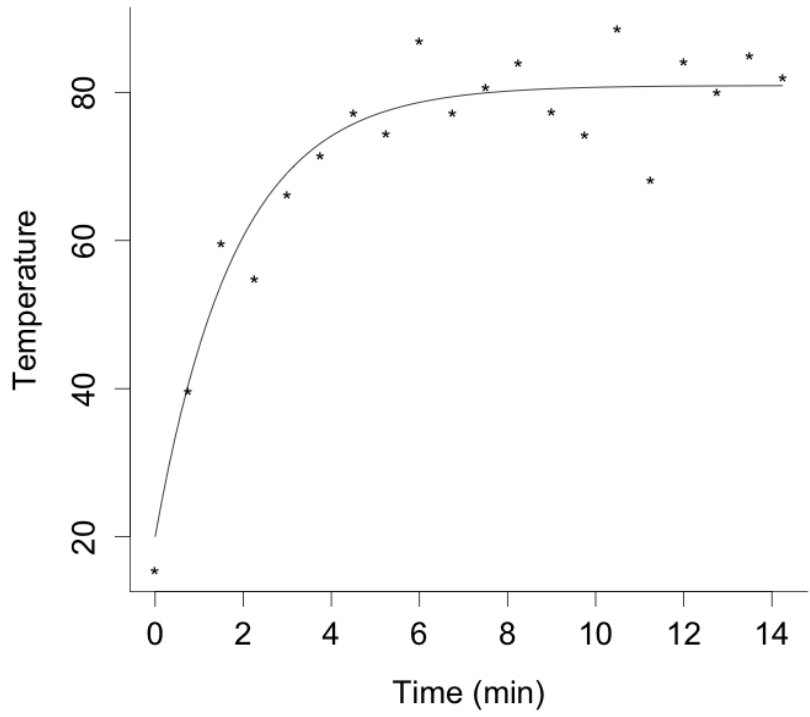}
		\caption{The solid line is the true temperature as a function of time from the Newton's law of cooling model with $x_1 =$ 20, $\theta = (-$0$.$5, 80$)^T$ and $n=20$. The scatter plot of the generated data of  temperatures and times is also drawn.}
		\label{fig:NLCm}
\end{figure*}

The model parameters were fixed at  $x_1 = 20, \theta = (-0.5, 80)^T$ and $\sigma^2 = 25$, and $y(t_i)$ were generated at $t_i =   h (i-1)$ for $i=1,2,\ldots, n$.
We generated 4 data sets with sample sizes $n=20, 50, 100, 150$, which have step sizes $h=0.75, 0.3, 0.15, 0.1$, respectively.
The effect of sample size on the approximation is investigated below. The data set with sample size $n=20$ and the true mean function is given in Figure \ref{fig:NLCm}.

The priors were set by
\begin{align}\label{NLCprior}
\begin{split}
x_1 \mid \tau^2 & \sim N(\mu_{x_1} =  y_1, 100/\tau^2) \\
\tau^2 & \sim Gamma(a, b) \\
\theta =(\theta_1, \theta_2)  & \sim Uniform(-200, 0)\times  Uniform(-200, 500).
\end{split}
\end{align}
where $a=0.1, b=0.01$ and $y_1=15.515$.

The true posterior of $\theta$ and $\tau^2$ can be obtained as follows:
\begin{eqnarray*}\label{truePost}
\tau^2 \mid \theta, {\bf y}_n &\sim& Gamma(\frac{np}{2}+a, \frac{1}{2}\tilde{u}(\theta) + b)\\
\theta \mid {\bf y}_n &\sim& \frac{1}{(\frac{1}{2}\tilde{u}(\theta)+b)^{\frac{np}{2}+a} } I(-200 < \theta_1 < 0)\\
&& \times  I(-200<\theta_2 < 500),
\end{eqnarray*}
where
\begin{eqnarray*}
\tilde{u}(\theta) &=& \mu_{x_1}^2/100 + \sum_{i=1}^n z_i^2 - (1/100 + \sum_{i=1}^n e^{2\theta_1(i-1)h})^{-1}\\
&& \times (\mu_{x_1}/100 + \sum_{i=1}^n z_ie^{\theta_1(i-1)h})^2, \\
z_i &=& z_i(\theta) =  y_i - \theta_2 + \theta_2 e^{\theta_1(i-1)h}.
\end{eqnarray*}

Since the dimension of $\theta$ is only $2$, the grid sampling is deemed to be adequate for sampling $\theta$.
For this example, we ended up setting  $M = 25$ and $h_0 = (1, 1)^T$ where $h_0$ is the vector of step sizes for grid matrix. The center of the grid matrix was chosen as $\theta^0 = (-0.547, 80.933)^T$ by parameter cascading method. In total, we have $2,601$ grid points. In the rest of the paper, we got 10,000 posterior sample from each example.

\subsubsection{Assessment of the performance of the approximate posteriors}\label{4.1.2}
The LAP inference has two approximations: Laplace approximation for the marginal posterior of $\theta$ and $\tau^2$ and numerical approximation method for the regression function $x$. With this example, we investigate the quality of these two approximations. In particular, we examine (1) the effect of sample size on the Laplace approximation and (2) that of  the numerical approximation. For the numerical approximation part, we compare the performance of the Euler method and the 4th order Runge-Kutta method.

\begin{figure*}[!hbpt]
\vspace{0cm}
\begin{tabular}{ccc}
		\includegraphics[width=0.35\textwidth]{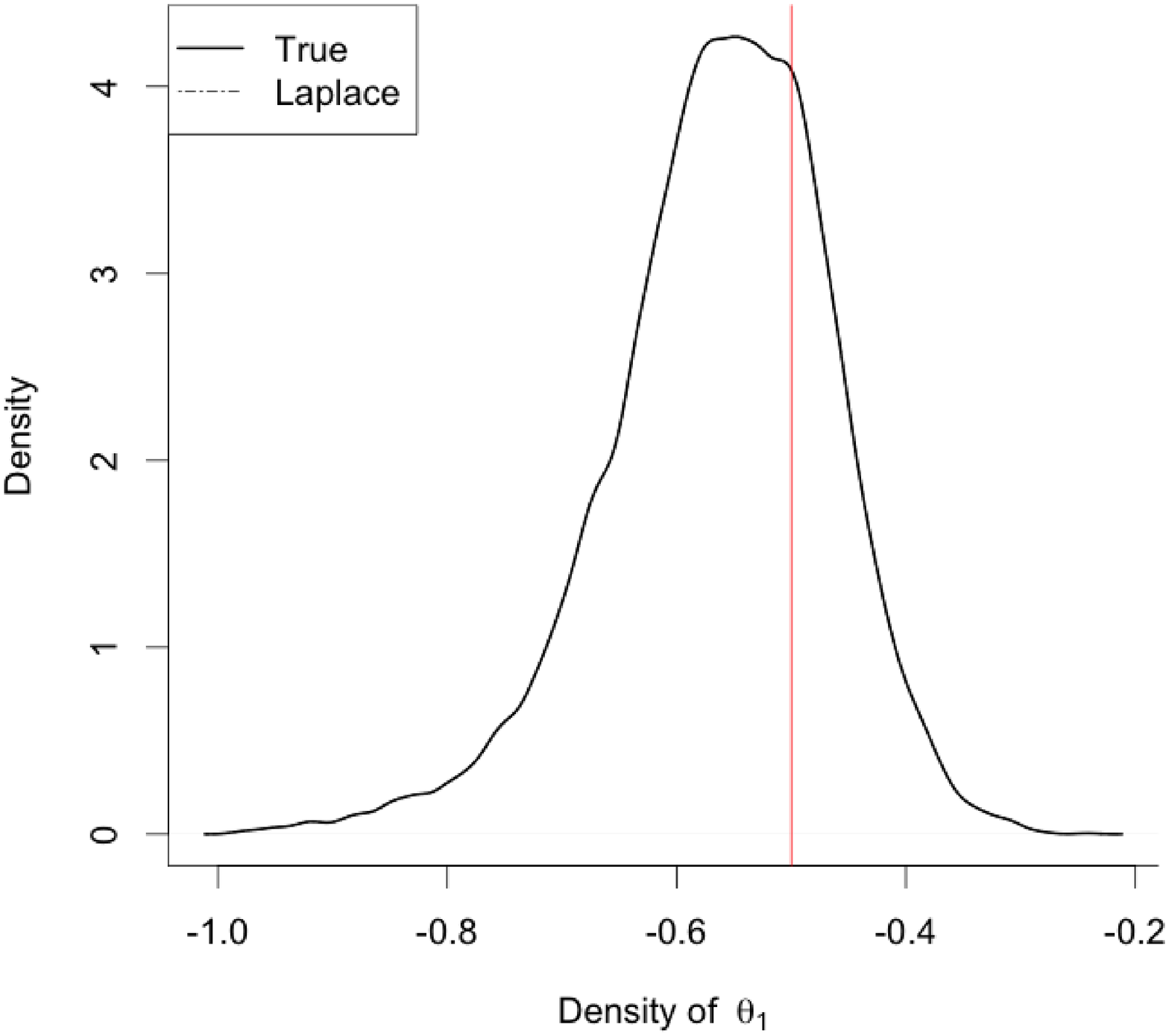} &\hspace{-1cm}
		\includegraphics[width=0.35\textwidth]{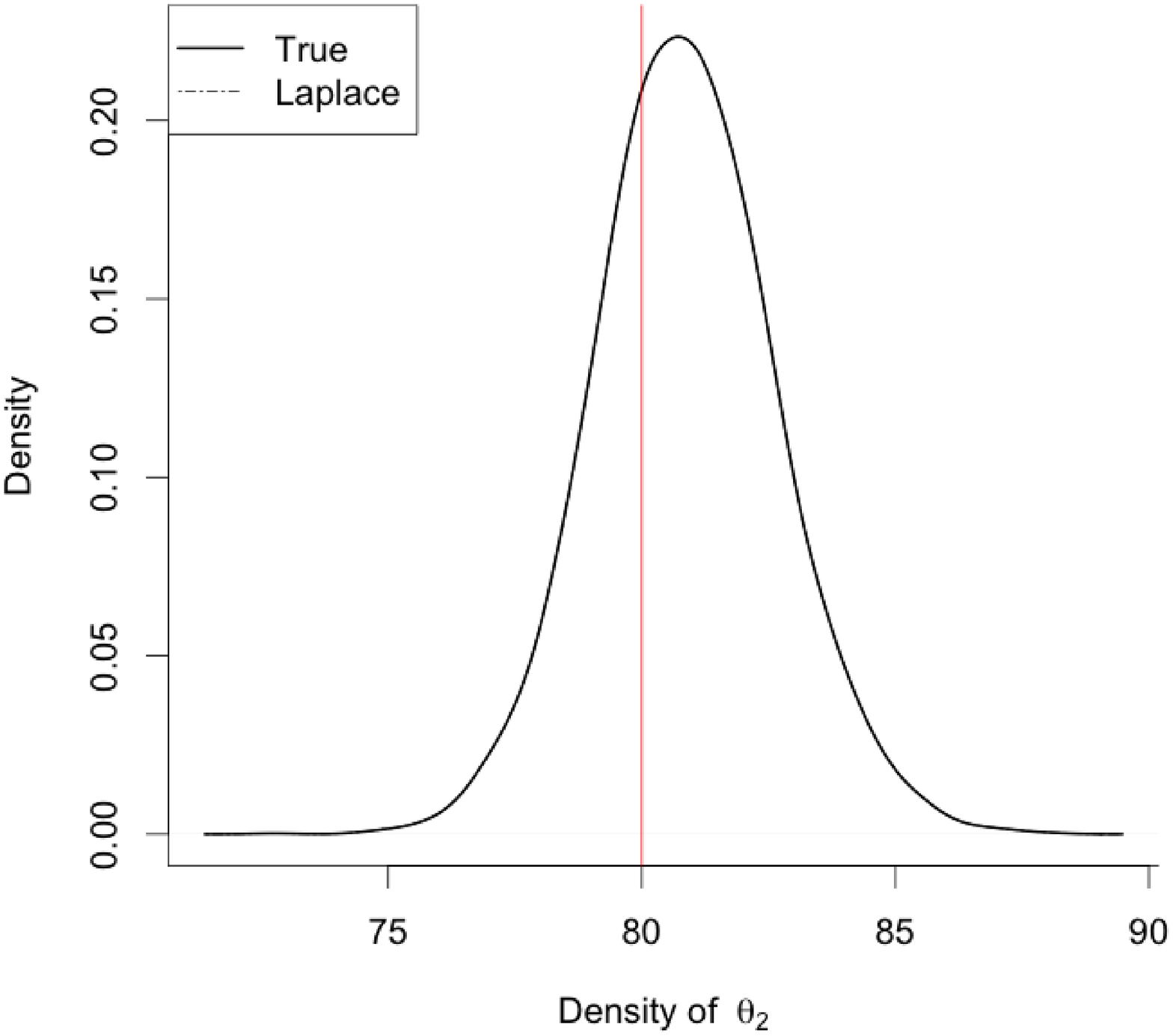} &\hspace{-1cm}
		\includegraphics[width=0.35\textwidth]{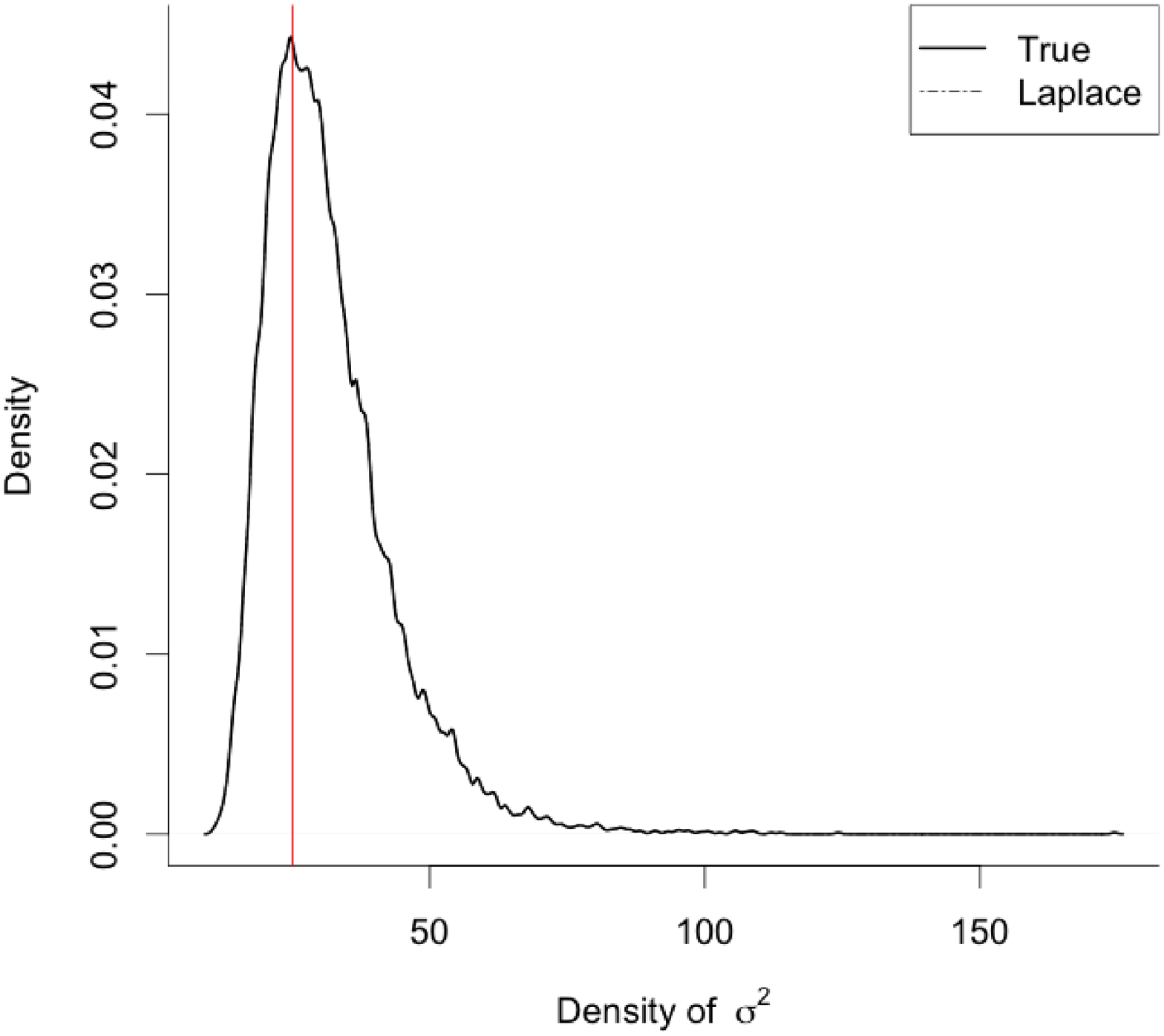}
\end{tabular}
\caption{The true posterior densities and the approximate posterior densities with Laplace approximation are shown.  The data are generated from Newton's law cooling model with sample size $n = 20$. The red lines represent true values of the parameters, $\theta = (-0.5, 80)^T$ and $\sigma^2 = 25$.}
		\label{fig:NLCdenLP}
\end{figure*}

To see the effect of sample size on the Laplace approximation, the true posterior $\pi(\theta, \tau^2 \mid {\bf y}_n)$ was compared with the posterior  with only Laplace approximation $\pi^{LP}(\theta, \tau^2 \mid {\bf y}_n)$. Figure \ref{fig:NLCdenLP} shows the true posterior densities and Laplace approximated posterior densities of each parameter when the sample size $n= 20$.
Even when the sample size is as small  as $n=20$, the Laplace approximated posterior densities are almost indistinguishable from the true posterior.
Although we have not shown here, we tried the same comparison plots for the samples with sample sizes as small as $5$ and $10$ and concluded that the approximation is still good.  Table \ref{table:NLC1} shows the similar story; that is,  the summary statistics of  the Laplace approximated posterior  are quite close to those of the true posterior. Table \ref{table:NLC1} shows only the summary statistics  of $\theta_1$, but the same conclusion has been reached for $\theta_2$ and $\tau^2$.

\begin{figure*}[!hbpt]
\begin{tabular}{ccc}
		\includegraphics[width=0.35\textwidth]{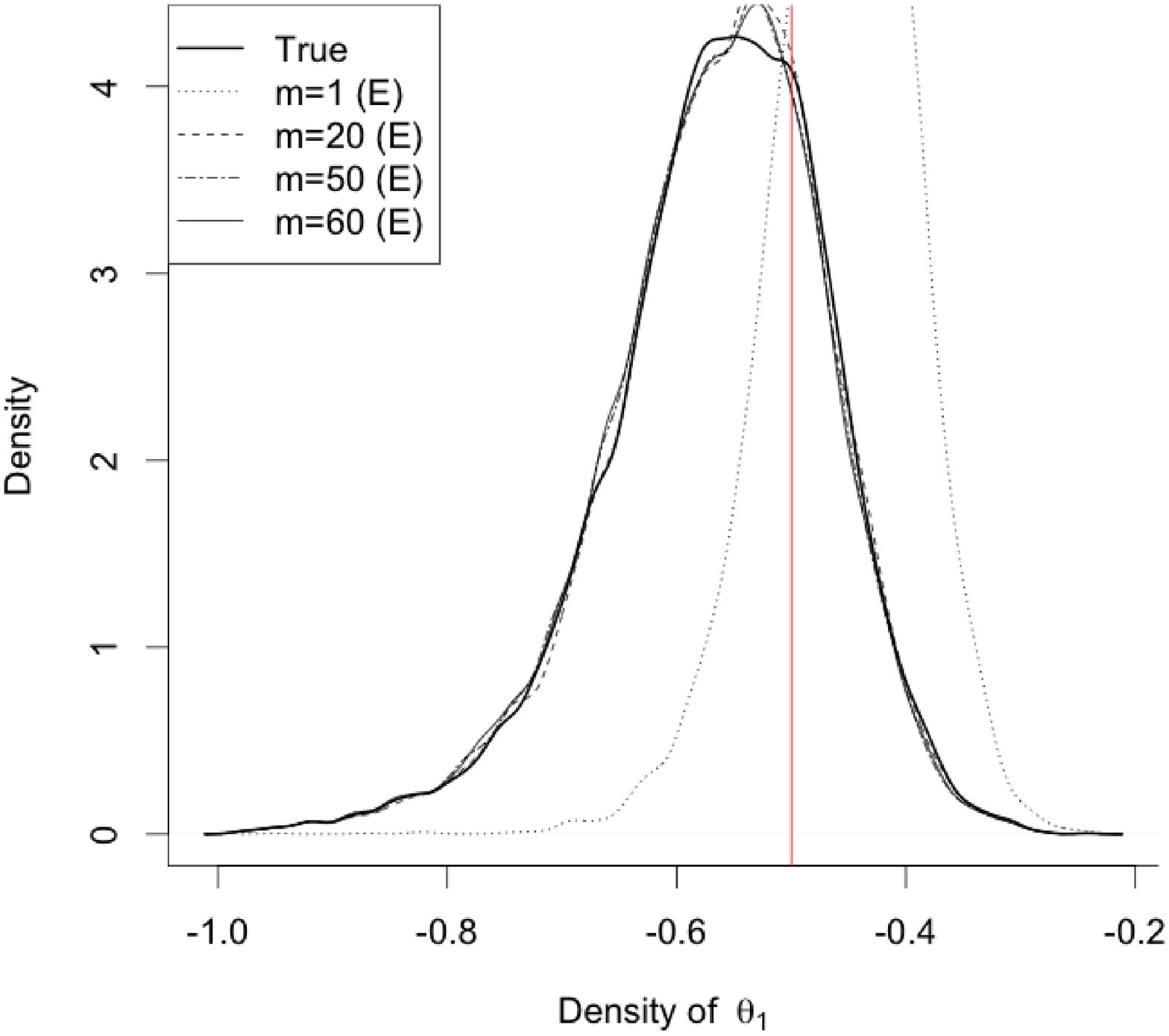} &\hspace{-1cm}
		\includegraphics[width=0.35\textwidth]{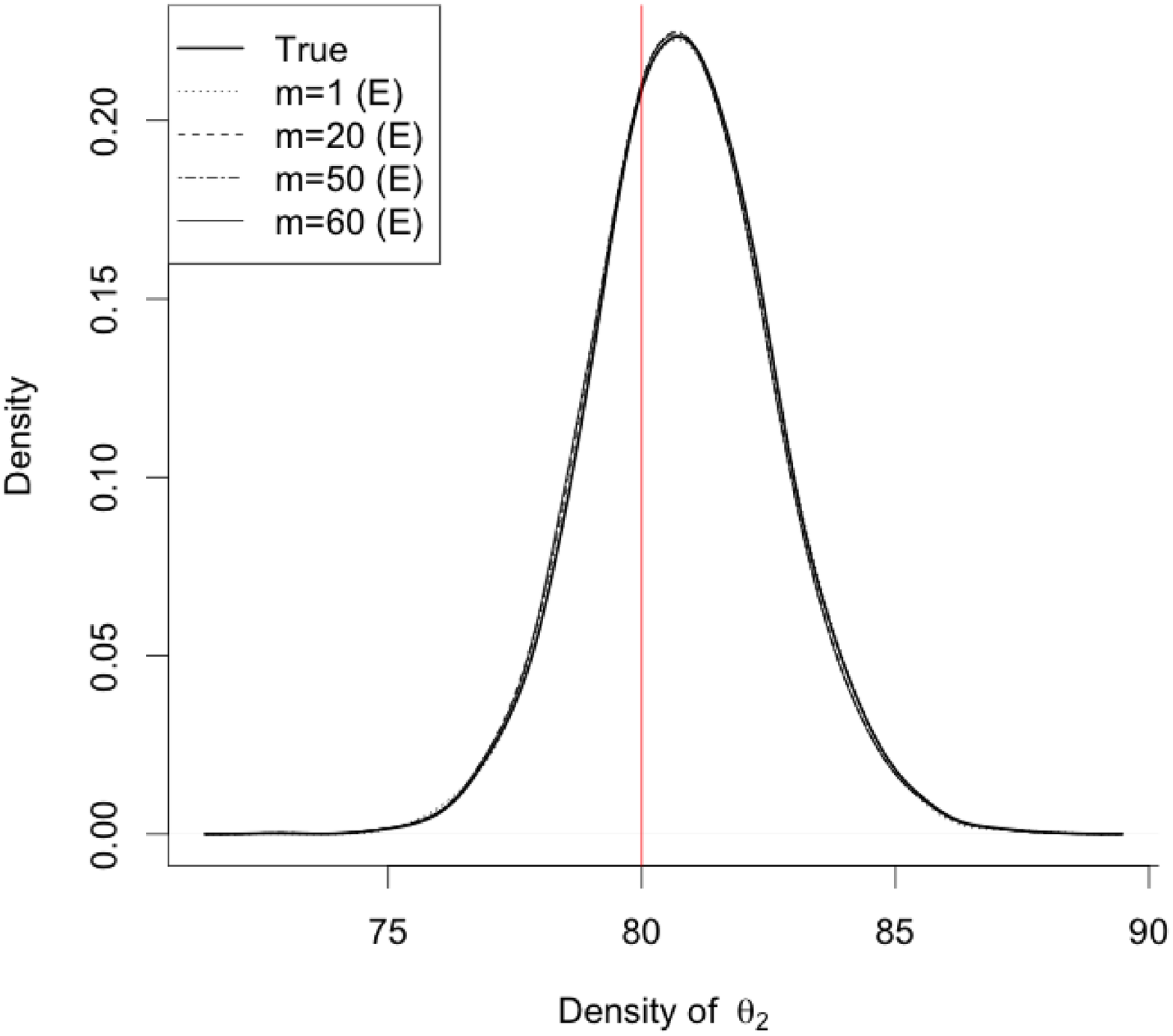} &\hspace{-1cm}
		\includegraphics[width=0.35\textwidth]{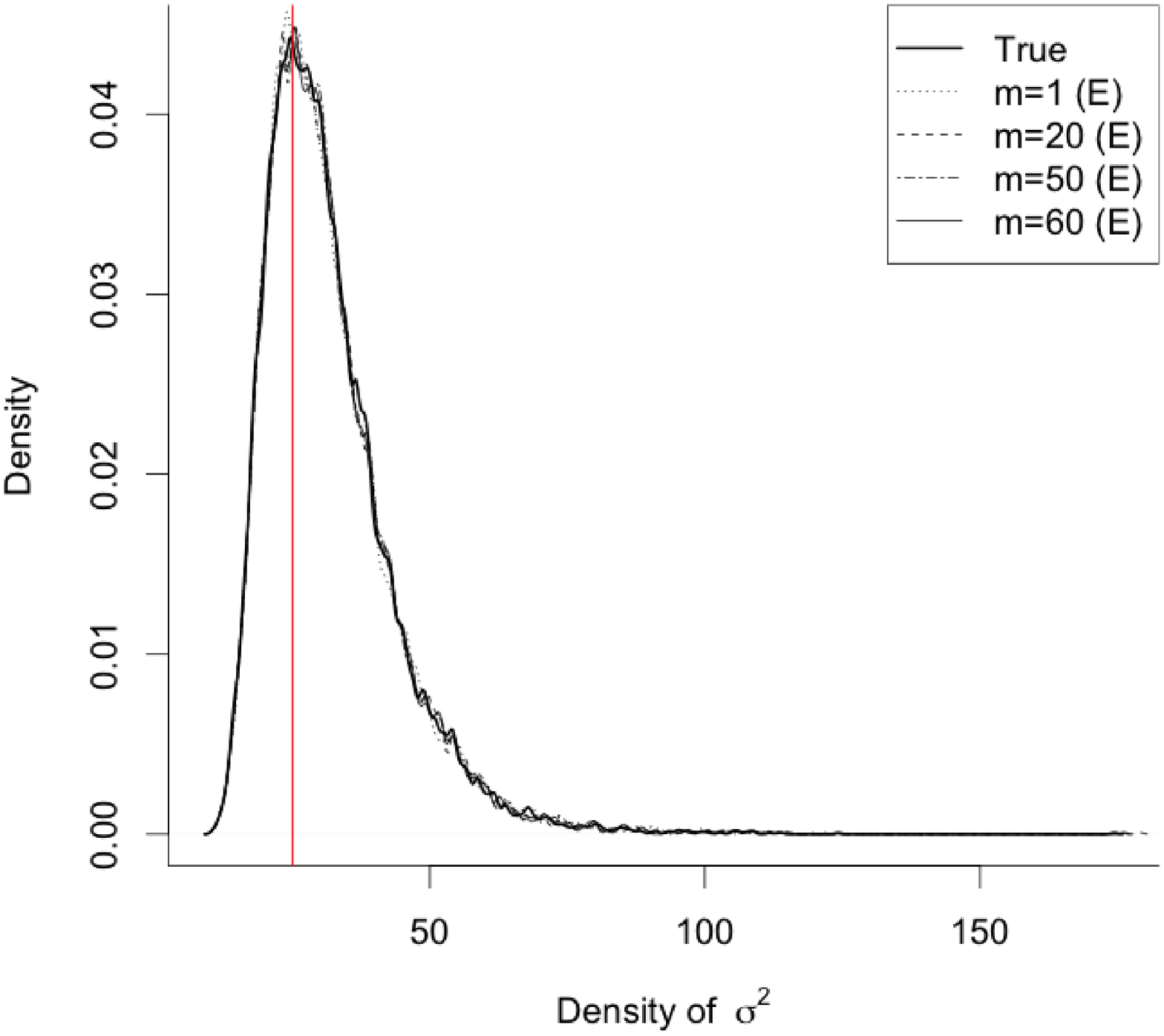} \\ \hspace{-1.5cm}
		\includegraphics[width=0.35\textwidth]{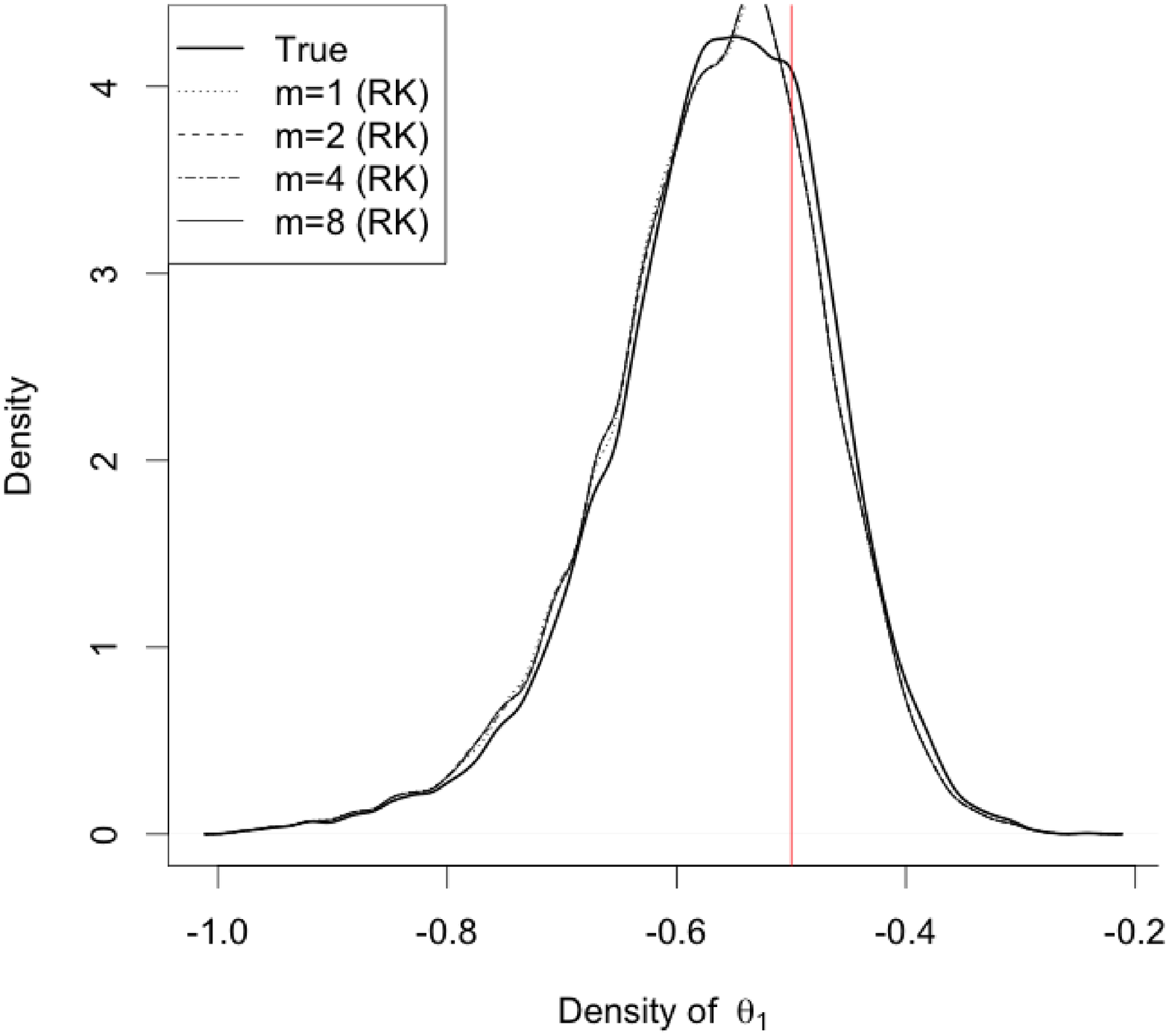} &\hspace{-1cm}
		\includegraphics[width=0.35\textwidth]{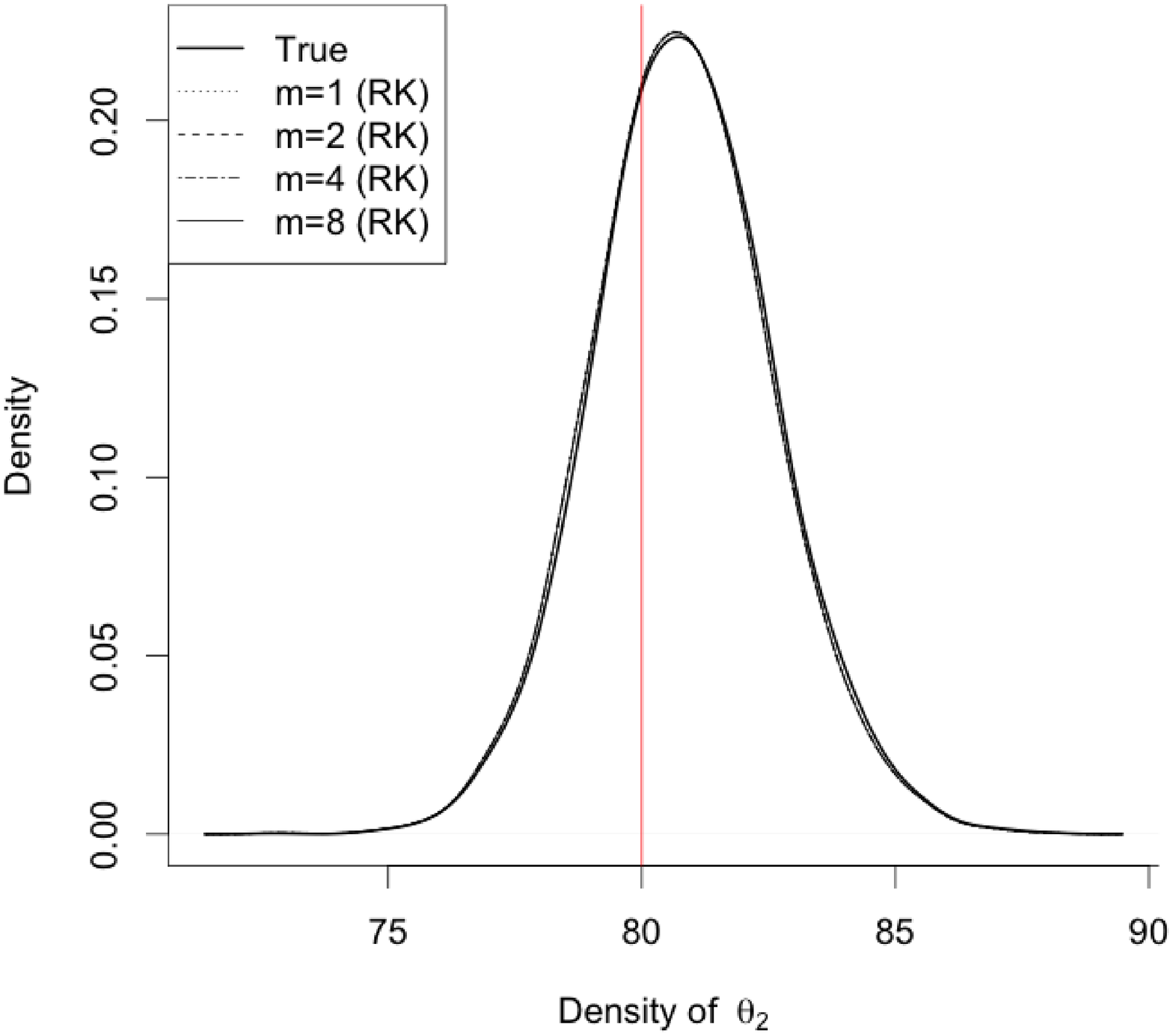} &\hspace{-1cm}
		\includegraphics[width=0.35\textwidth]{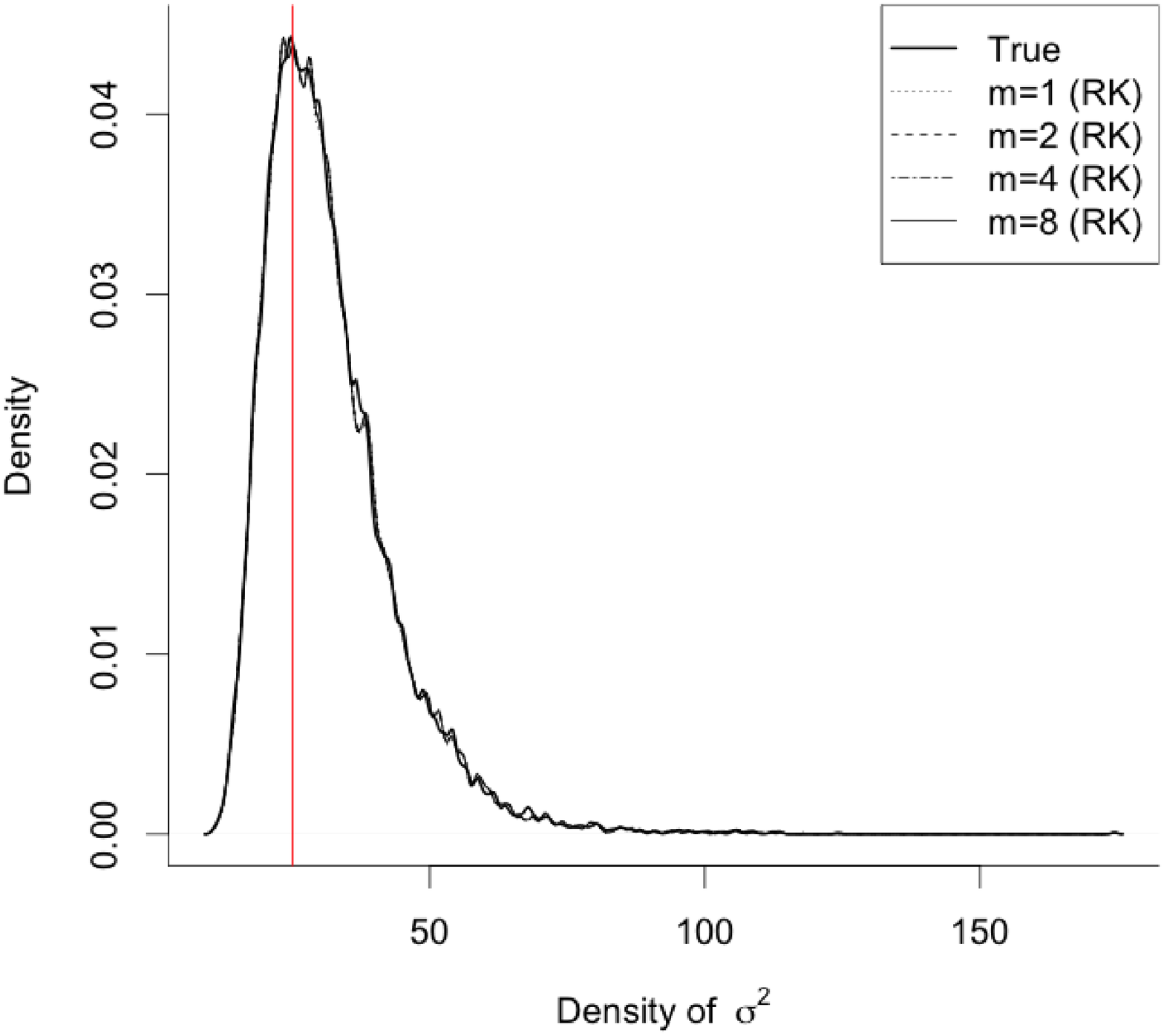}
\end{tabular}
\caption{The true posterior density and the approximate posteriors with Laplace approximation and numerical methods for the Newton's law of cooling model are drawn when $n = 20$. The true parameter values are $\theta = (-0.5, 80)^T$ and $\sigma^2 = 25$. As $m$ grows, the approximate posterior is getting closer to the true posterior.}
		\label{fig:NLCden}
\end{figure*}

To see the effect of the approximation due to the numerical methods, the true posterior $\pi(\theta, \tau^2 \mid {\bf y}_n)$ was compared with the approximate posteriors obtained by applying the Laplace and the numerical methods. In this example, we used the Euler method and 4th order Runge-Kutta method for the numerical method. The intervals between observations $[t_{i-1}, t_i ]$ were divided into $m$ segments with $m=1,2, 4, 8, 14, 20, 30 , \ldots$. The approximate posteriors are denoted by $\pi^{LP,E}_m (\theta, \tau^2 \mid {\bf y}_n)$ and $\pi^{LP,RK}_m (\theta, \tau^2 \mid {\bf y}_n)$ where $E$ and $RK$ stand for the Euler and Runge-Kutta, and $m$ is the number of segments.
Figure \ref{fig:NLCden} shows the posterior densities with different $m$ and the true posterior density when the sample size $n = 20$.
The approximate posteriors $\pi^{LP,E}_m$  and $\pi^{LP,RK}_m$   are shown in the first row and the second row, respectively. The approximate posteriors $\pi^{LP,RK}_m$ are generally close to the true posterior even for $m=1$, but  $\pi^{LP,E}_m$  show different behavior. For $\theta_2$ and $\tau^2$, $\pi^{LP,E}_m$ are close to the true posterior even for $m=1$, but the marginal posterior of $\theta_1$ of $\pi^{LP,E}_m$ deviates from the true posterior. The deviation disappears as $m$ gets larger. These results can be also confirmed in Table \ref{table:NLC1} which includes the posterior summary statistics of $\theta_1$ with different values of $m$ and $n$.
We represent the results for the Euler with $m=1,20,50,60$ and the Runge-Kutta with $m=1,2$.
Based on these observations, we recommend the Euler with $m=50$ and the Runge-Kutta with $m=1$.
In Sect. \ref{sec:4}, we present a theorem, a theoretical basis for this observation. The Runge-Kutta method with $m=1$ does not reduce the error rate obtained by Laplace method, while Euler with $m =1$ does reduce the error rate and the larger value of $m$ is needed for the Euler method.
The computation times for the numerical methods with various values of $m$ and $n$ in this example are shown in Table \ref{table:NLCtime}.

\begin{table}[!t]
\scriptsize\centering
\caption{Posterior summary statistics of $\theta_1$ from the true posterior, Laplace approximated posterior, posterior with Laplace approximation and numerical approximation method with varying values of the number of steps $m$ and sample sizes $n$ in Newton's law of cooling model.}
\begin{tabular}{| c | c | c | c | c | c |}
\hline
\multicolumn{3}{|c|}{}  &  \multicolumn{3}{| c |}{$\theta_1$}  \\ \hline
n & \multicolumn{2}{|c|}{Case} & Mean & Median & 90\% credible interval \\ \hline
\multicolumn{1}{|c|}{\multirow{8}{*}{n = 20}} &  \multicolumn{2}{|c|}{$\pi$}  &  -0.563 & -0.555 & (-0.734, -0.421)  \\ \cline{2-3}
 & \multicolumn{2}{|c|}{$\pi^{LP}$}  &  -0.563 & -0.555 & (-0.734, -0.421)   \\ \cline{2-3}
 & \multicolumn{1}{|c|}{\multirow{4}{*}{$\pi^{LP,E}$}} & m=1  & -0.457 & -0.453 & (-0.565, -0.360)  \\ \cline{3-3}
& & m=20  & -0.563 & -0.553 & (-0.736, -0.423)  \\ \cline{3-3}
& & m=50 & -0.567 & -0.557 & (-0.740, -0.425)  \\ \cline{3-3}
& & m=60  & -0.567 & -0.559 & (-0.742, -0.425)  \\ \cline{2-3}
& \multicolumn{1}{|c|}{\multirow{2}{*}{$\pi^{LP,RK}$}} & m=1  & -0.569 & -0.561 & (-0.744, -0.427)  \\ \cline{3-3}
& & m=2  & -0.569 & -0.561 & (-0.744, -0.427)  \\ \hline
 \multicolumn{1}{|c|}{\multirow{8}{*}{n = 50}} &  \multicolumn{2}{|c|}{$\pi$}  &  -0.589 & -0.585  & (-0.711, -0.482)  \\ \cline{2-3}
 & \multicolumn{2}{|c|}{$\pi^{LP}$}  &  -0.589 & -0.585  & (-0.711, -0.482)  \\ \cline{2-3}
 & \multicolumn{1}{|c|}{\multirow{4}{*}{$\pi^{LP,E}$}} & m=1  & -0.581 & -0.581 & (-0.585, -0.576)  \\ \cline{3-3}
& & m=20  & -0.589 & -0.585 & (-0.711, -0.482)  \\ \cline{3-3}
& & m=50 & -0.591 & -0.586 & (-0.711, -0.482)  \\ \cline{3-3}
& & m=60  & -0.591 & -0.587 & (-0.711, -0.482)  \\ \cline{2-3}
 & \multicolumn{1}{|c|}{\multirow{2}{*}{$\pi^{LP,RK}$}} & m=1  & -0.592 & -0.588 & (-0.711, -0.482)  \\ \cline{3-3}
& & m=2  & -0.591 & -0.588 & (-0.711, -0.482)  \\ \hline
\end{tabular}
\label{table:NLC1}
\end{table}

\begin{table}[htpb]
\scriptsize\centering
\caption{The computation times (s) for numerical methods with varying values of step size $m$ and sample sizes in numerical approximation method for Newton's law of cooling model.}
\begin{tabular}{|c|c|c|c|c|}\hline
n & m & Euler & m &  4th order Runge-Kutta \\ \hline
\multirow{4}{*}{20} & 1 & 0.370 & \multirow{2}{*}{1}  & \multirow{2}{*}{1.107} \\  \cline{2-3}
 & 20 & 1.809 & & \\ \cline{2-5}
 & 50 & 4.095 & \multirow{2}{*}{2} & \multirow{2}{*}{1.750} \\ \cline{2-3}
 & 60 & 4.663 & & \\ \hline
\multirow{4}{*}{50} & 1 & 4.955 & \multirow{2}{*}{1}  & \multirow{2}{*}{2.535} \\ \cline{2-3}
& 20 & 4.178 & & \\ \cline{2-5}
& 50 & 9.243 & \multirow{2}{*}{2} & \multirow{2}{*}{3.821} \\ \cline{2-3}
& 60 & 10.981 & & \\ \hline
\multirow{4}{*}{100} & 1 & 1.454 & \multirow{2}{*}{1}  & \multirow{2}{*}{4.618} \\  \cline{2-3}
 & 20 & 8.015 & & \\ \cline{2-5}
 & 50 & 18.138 & \multirow{2}{*}{2}  & \multirow{2}{*}{7.192} \\ \cline{2-3}
 & 60 & 21.793 & & \\ \hline
\multirow{4}{*}{150} & 1 & 1.992 & \multirow{2}{*}{1}  & \multirow{2}{*}{6.452} \\ \cline{2-3}
& 20 & 11.315 & & \\  \cline{2-5}
& 50 & 25.936 & \multirow{2}{*}{2}  & \multirow{2}{*}{9.944} \\ \cline{2-3}
& 60 & 30.459 & & \\ \hline
\end{tabular}\label{table:NLCtime}
\end{table}

\subsection{FitzHugh-Nagumo model}\label{FN}
\subsubsection{Model description and data generation}
The action of spike potential in the giant axon of squid neurons is modeled by Hodgkin and Huxley (1952)\nocite{Hodgkin52}.
FitzHugh (1961)\nocite{Fithugh61} and Nagumo et al. (1962)\nocite{Nagumo62} simplified this model with two variables. The reduced model with no external stimulus is given below:
\begin{eqnarray*}\label{GS}
{\dot x_1}(t) & = & \theta_3 (x_1(t) - \frac{1}{3}x_1^3(t) + x_2(t)),\nonumber \\
{\dot x_2}(t) & = & -\frac{1}{\theta_3}(x_1(t) - \theta_1 + \theta_2 x_2(t)),
\end{eqnarray*}
where $-0.8 < \theta_1, \theta_2 < 0.8, 0 < \theta_3 < 8$,  and $x_1(t)$ and $x_2(t)$ are the voltage across an membrane and outward currents at time $t$ and called the voltage and recovery variables, respectively.
We use this parameter space for stable cyclical behavior of the system (Campbell 2007).
With this example, we show that the Laplace approximated posterior inference works well with appropriate choice of $m$.

We generated a simulated data set with model parameters $\theta = (0.2, 0.2, 3)^T, x_1 = x(t_1) = (-1, 1)^T$ and $\sigma^2 = 0.25$.
The time interval was fixed at $t_i - t_{i-1} = 0.2$ for $i=2, 3, \ldots, n$ with $n=100$. We divided $[t_{i-1}, t_i]$ into 100 segments and applied the 4th order Runge-Kutta method to got the true mean function for simulated data.

For the prior, we had $x_1 \mid \tau^2 \sim N_2( \mu_{x_1} = y_1, 100/\tau^2 I_2)$, $\tau^2 \sim Gamma(a, b)$ and $\theta \sim Unif(A)$ where $a=0.1, b=0.01$, $y_1 = (-1.449, 1.092)^T$ and
$A = \{(\theta_1, \theta_2, \theta_3) : -0.8 < \theta_1,\theta_2 < 0.8, 0 < \theta_3 < 8 \}$.

\subsubsection{Assessment of the performance of the approximate posteriors}
\begin{figure*}[!htpb]	
	\centering
		\includegraphics[width=.45\textwidth]{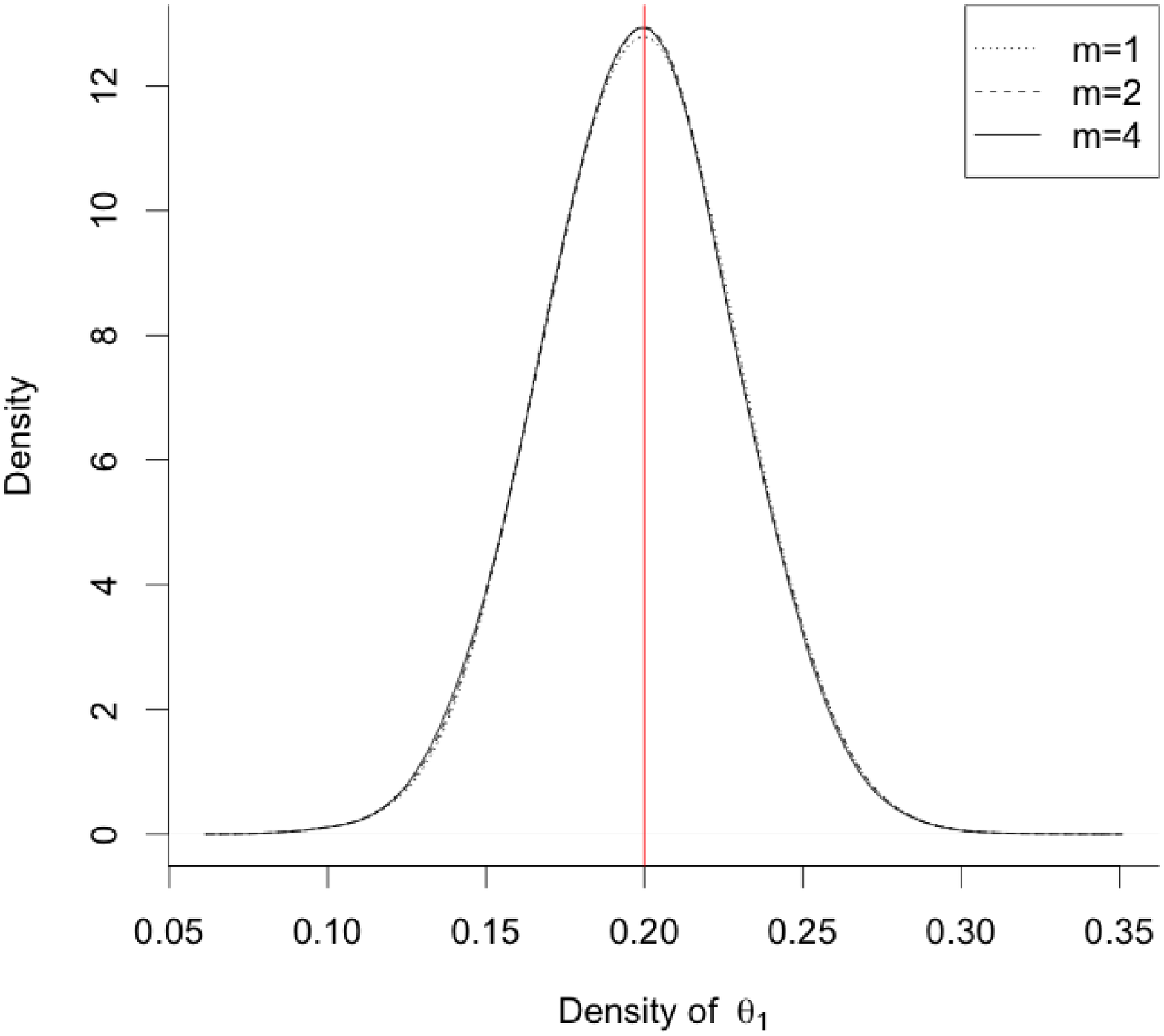}
		\includegraphics[width=.45\textwidth]{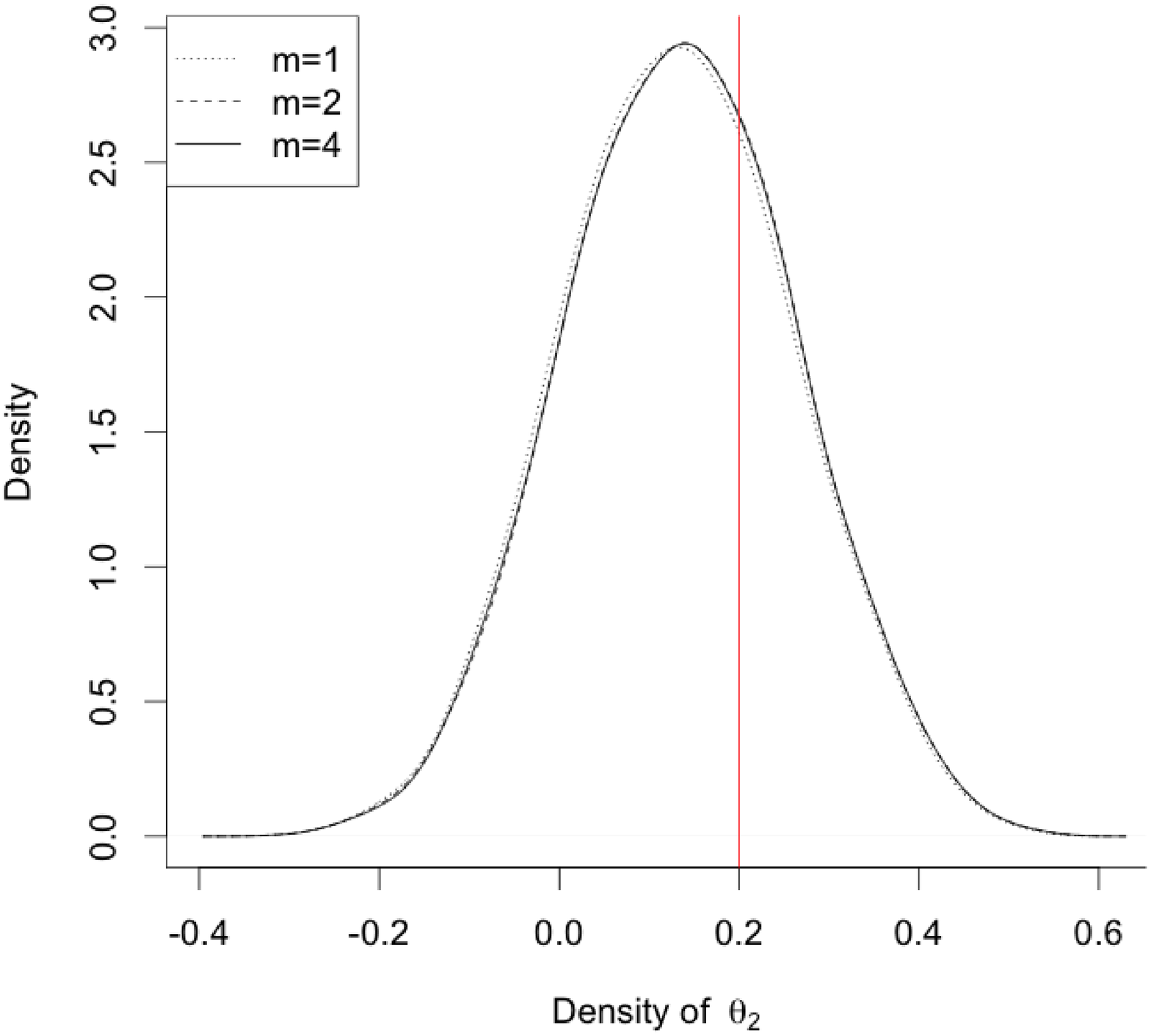}
	  	\includegraphics[width=.45\textwidth]{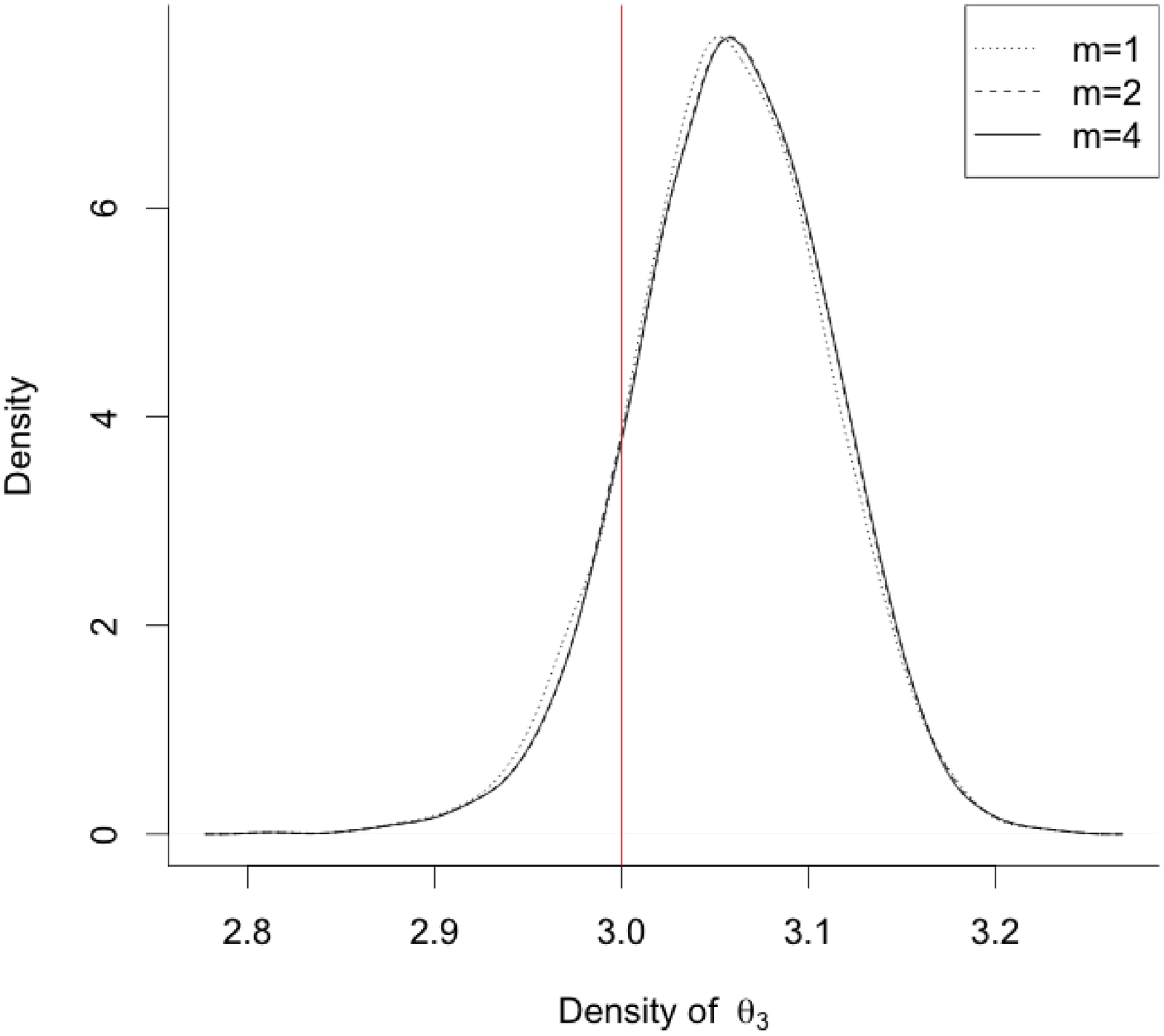}
		\includegraphics[width=.45\textwidth]{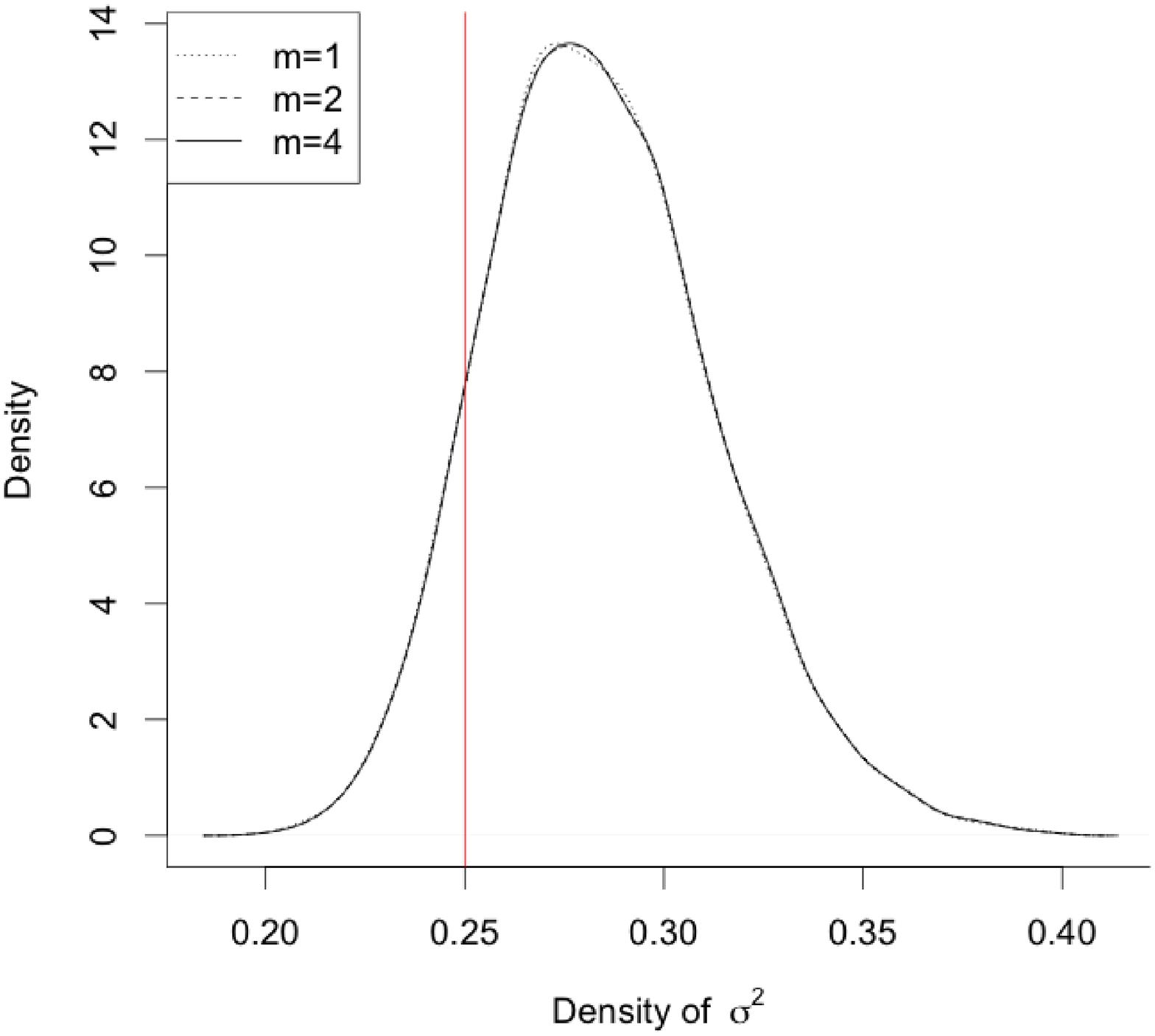}
		\caption{Approximate marginal posterior densities for each parameter with varying values of $m$ for FitzHugh-Nagumo model. The red lines represent true values of the parameters, $\theta = (0.2,0.2,3)^T$ and $\sigma^2 = 0.25$. As $m$ grows, the approximate posterior seems to be stabilized.}
	\label{fig:GSden}			
\end{figure*}

We applied the procedure in Sect. \ref{sec:3}, to choose the range and the center of the grid matrix. For the final analysis, we set  $M = 15$ and $h_0 = (7, 7, 4)^T$, so we have $29,791$ grid points. The center of the grid matrix $\theta^0$ was $(0.199, 0.131, 3.056)^T$.
Figure \ref{fig:GSden} shows the posterior densities with different $m$.
In this example, different values of $m$ shows slight changes in the posterior approximations.
Table \ref{table:GS1} shows the posterior summary statistics with varying values of step size.
We applied the procedure to choose $m$ described in Sect. \ref{sec:3} and in the final analysis $m=2$ was used, and  Table \ref{table:GStime} contains the computation times for $m=1,2$ and $n=100, 200$.

\begin{table}[!tb]
\caption{Posterior summary statistics with varying values of step size $m$ in 4th order Runge-Kutta method for FitzHugh-Nagumo model. In the table, C.I. denotes the credible interval.}
\scriptsize
\begin{tabular}{|c|c|c|c|c|c|c|}
  \hline
  & \multicolumn{3}{c|}{$\theta_1$} & \multicolumn{3}{|c|}{$\theta_2$} \\ \hline
  m & Mean & Median & 90\% C.I. & Mean & Median & 90\% C.I. \\ \hline
  1 & 0.199 & 0.190 & (0.150, 0.248) & 0.130 & 0.132 & (-0.074, 0.350)  \\ 
  2 & 0.198 & 0.198 & (0.150, 0.247) & 0.135 & 0.134 & (-0.071, 0.352)  \\ 
  4 & 0.198 & 0.198 & (0.149, 0.246) & 0.135 & 0.134 & (-0.070, 0.352)  \\ \hline
& \multicolumn{3}{c|}{$\theta_3$} & \multicolumn{3}{|c|}{$\sigma^2$} \\ \hline
m & Mean & Median & 90\% C.I. & Mean & Median & 90\% C.I. \\ \hline
1 & 3.057 & 3.057 & (2.968, 3.143) & 0.284 & 0.282 & (0.241, 0.335) \\ 
2 & 3.059 & 3.061 & (2.972, 3.143) & 0.285 & 0.283 & (0.241, 0.335) \\ 
4 & 3.060 & 3.061 & (2.972, 3.143) & 0.285 & 0.282 & (0.241, 0.335) \\ \hline
\end{tabular}
\label{table:GS1}
\end{table}

\begin{table}[!tb]
\scriptsize\centering\vspace{0cm}
\caption{The computation times (s) with varying values of step size $m$ and sample size in numerical approximation method for the FitzHugh-Nagumo model.}
\begin{tabular}{|c|c|c|}\hline
n & m & 4th order Runge-Kutta  \\ \hline
\multirow{2}{*}{100} & 1 & 111.137   \\
& 2 & 172.215   \\ \hline
\multirow{2}{*}{200} & 1 & 131.46   \\
& 2 & 200.973  \\ \hline
\end{tabular}\label{table:GStime}
\end{table}

\begin{figure*}[!t]
		\includegraphics[width=.5\textwidth]{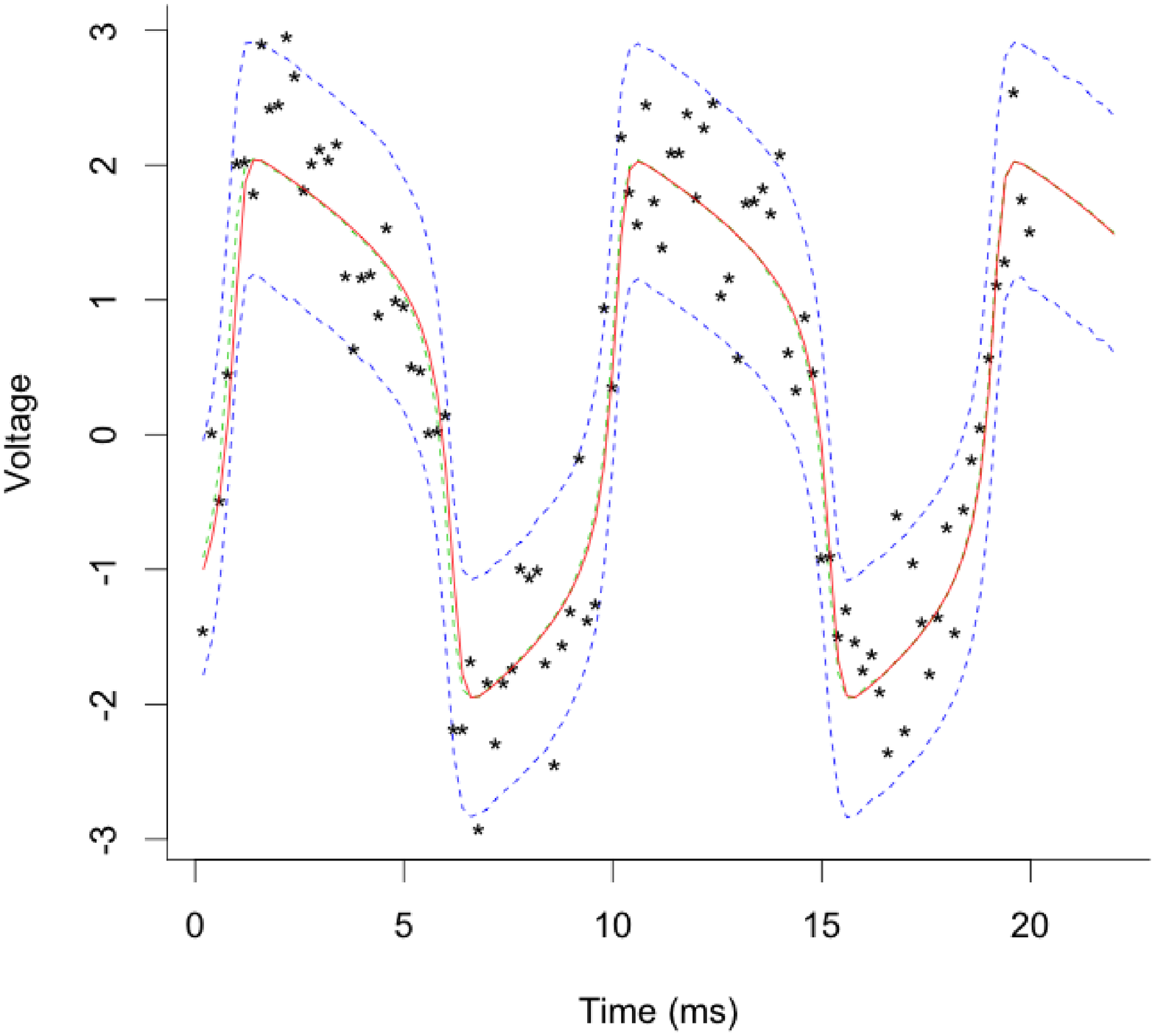}
		\includegraphics[width=.5\textwidth]{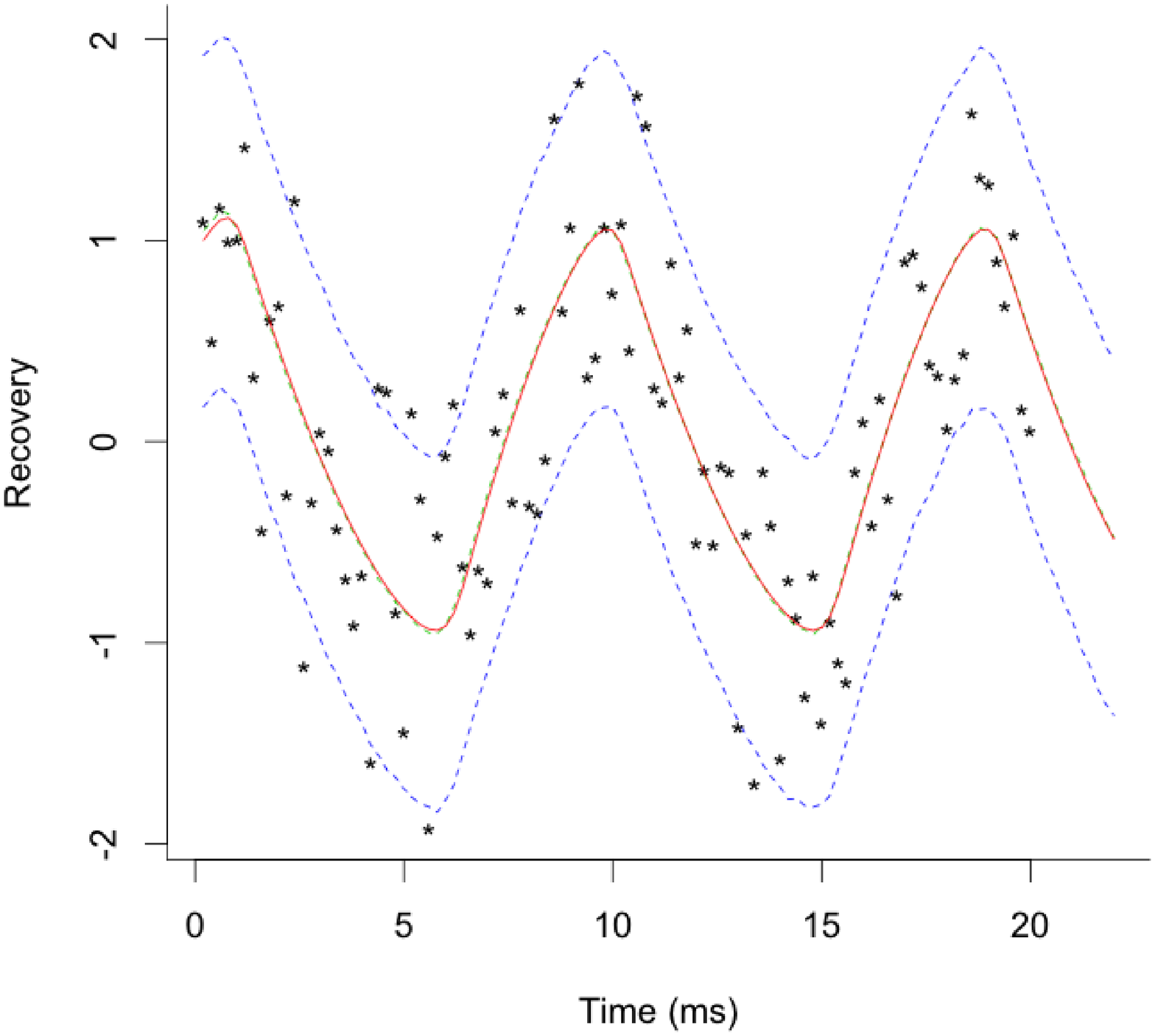}\vspace{-0.5cm}
		\caption{Scatter plot of the observations generated from the FitzHugh-Nagumo model, and plots of $90\%$ credible set lines and true states are drawn when $m=2$. Predictions of 10 time points ahead are also drawn. The upper, lower and middle dotted lines are the 95\% and 5\% quantiles  and mean of the posterior, respectively. The solid line in the middle is the true value of the state $x(t)$, and the star-shaped points are the observations.}
		\label{fig:GS1pred}
\end{figure*}
Figure \ref{fig:GS1pred} contains the scatter plots of the observations, the true mean functions, $90\%$ credible lines for the mean functions, and the posterior mean functions as well as prediction values at $10$ future time points when $m =2$.

\subsubsection{Comparison with existing methods}
We compare the performance of the LAP inference and the other existing methods:  the parameter cascading method  (Ramsay et al. 2007), the delayed rejection adaptive Metropolis (DRAM) algorithm (Haario et al. 2006)\nocite{haario06} with numerical integration, the Gaussian Process-ODE (GP-ODE) approach (Wang and Barber 2014)\nocite{barber14} and the adaptive gradient matching (AGM) approach (Dondelinger et al. 2013)\nocite{dondelinger2013ode}.
We generate 100 simulated data set as above and compute the absolute bias, the standard deviation, the root mean squared error (rmse) and the log-likelihood to use as the measure of performance.
However, for the GP-ODE approach (Wang and Barber 2014) and AGM approach (Dondelinger et al. 2013), only 20 data set were used because of their long computation times. This long computation times is mainly due to the fact that the implementations of the two approaches were based on the pure MATLAB codes. 
For the data generation, $\theta=(0.2, 0.2, 3)^T, x_1=(-1,1)^T,$ $\sigma^2=0.25, h=0.2$ and $n=30$ were used.

We are using the method of Ramsay et al. (2007) based on parameter cascading (PC). PC is also called generalized profiling. We give the details below.

To represent the state of the ODE, $x(t)$, the PC methodology uses the collocation method: The collocation method uses a series of basis expansion to represent the $p$ dimensional vector $x(t)$, that is, 
\begin{equation}
\label{basisexpansion}
x(t) =(x_1(t),x_2(t),\cdots,x_{p}(t)) = \Phi(t)\mathbf{C}
\end{equation}
 where $\Phi(t) = (\Phi_1(t), \cdots , \Phi_K(t))$ is a set of $K$ bases evaluated at time $t$, and the $K \times p$ matrix $\mathbf{C}$ contains the coefficients of the basis functions of each variable in its columns. In other words, expanding (\ref{basisexpansion}), the $i$-th component of $x(t)$ at time $t$ has the basis function expansion
\begin{equation}
\label{basisexpansioni}
x_i(t) = \Phi(t)c_i = \sum_{k=1}^{K}\,c_{ik}\Phi_{k}(t),
\end{equation}
where $c_i$ is a column vector of coefficients $c_{ik}$ of length $K$, for $i=1,2,\cdots,p$. The ODE model whose parameters need to be estimated is given by \begin{equation}
\label{ODEmodel}\dot{x}_{i}(t)=f_{i}(x(t),{\theta})\end{equation} for $i=1,2,\cdots,p$.

PC involves a penalized likelihood criteria $J\equiv J(\mathbf{C},\theta,\boldsymbol{\lambda})$ which is based on the coefficients of basis expansions $\mathbf{C}$, the unknown parameter $\theta$ to be estimated and $\boldsymbol{\lambda}\equiv (\lambda_1, \cdots , \lambda_p)$, the penalty (or smoothing) parameters.  The criteria $J$ is reflects two competing goals based on two competing terms. The first term in $J(\mathbf{C},\theta,|\boldsymbol{\lambda})$ measures how well the state function values fit the data whereas the second term measures how closely each of the state functions satisfy the corresponding differential equation (\ref{ODEmodel}). The smoothing parameters measures the weight of each competing term; when $\lambda_i$s, $i=1,2,\cdots,p$, are large, more and more emphasis is put on having $x_i(t)$s in (\ref{basisexpansioni}) satisfy the differential equation in (\ref{ODEmodel}), as opposed to fitting the data and vice versa when $\lambda_i$s tend to zero.

PC optimization is based on two levels: An inner optimization step nested within an outer optimization. In the inner optimization, ${\theta}$ and $\boldsymbol{\lambda}$ components are held fixed, and an inner optimization criterion is optimized with respect to the coefficients in matrix $\mathbf{C}$ only. In effect, this makes $J = J(\mathbf{C}(\theta, \lambda),\theta,\boldsymbol{\lambda})$ a function of ${\theta}$ and $\boldsymbol{\lambda}$ only. In the outer optimization step, $J$ is optimized with respect to $\theta$ keeping $\boldsymbol{\lambda}$ fixed. This essentially makes $J \equiv J(\mathbf{C}(\theta(\boldsymbol{\lambda}),\lambda),\theta(\boldsymbol{\lambda}),\boldsymbol{\lambda})$ now a function of $\boldsymbol{\lambda}$ only. The smoothing parameters $\boldsymbol{\lambda}$ and number of basis functions $K$ are finally chosen based on numerical stability of the parameter estimates. This is the key idea underlying the generalized profiling or parameter cascade algorithms in Ramsay (2007) and Cao and Ramsay (2009)\nocite{cao09}.

The implementation of the PC method was carried out using the \verb2CollocInfer2 package (Hooker et al. 2014)\nocite{Hooker2014collocinfer} in R. This package uses B-spline basis functions for $\Phi_{k}(t), k = 1, . . . , K$. B-spline basis functions are constructed by joining polynomial segments end-to-end at junctions specified by knots.
Since our method used $m=2$, we set $2n-1$ equally spaced knots on $[t_1, t_n]$ to get a twice number of knots than the data points. 
The finer knots gave negligible improvement in parameter estimate while slowing down the computational speed. 
We chose the three-order of B-spline basis which was used in Ramsay et al. (2007) for the same model.
For the choice of the tuning parameter $\lambda$, we used both the manual procedure and the automatic procedure. 
We adopted the procedure of Ramsay et al. (2007) which tries larger values of $\lambda$ and chooses $\lambda$ manually which gives a stable result.
The quartiles of the parameter estimates for 100 simulation data sets were obtained as $\lambda$ is varied from $10^{-2}$ to $10^6$.
After that, this $\lambda$ set at $10^5$.
For the automatic procedure, the forward prediction error (FPE) in Hooker et al. (2010)\nocite{hooker10} was used. 
We divided the each data set into ten part, from $t_1$ to $t_{10}$, and $t_{11}$ to $t_{20}$, and so on. For one data set, the averaged FPE was obtained as $\lambda$ is varied from $10^{-2}$ to $10^6$, and the optimal $\lambda$ which minimizes FPE was chosen.

The DRAM algorithm (Haario et al. 2006) is a variant of the standard Metoropolis-Hastings algorithm (Metropolis et al. 1953; Hastings 1970).\nocite{metropolis53}\nocite{hastings70}
We chose the DRAM algorithm with numerical integration to compare the computation time with our LAP inference.
To implement it, we used the \verb|modMCMC| function from the \verb|FME| package\nocite{Soetaert10package} (Soetaert and Petzoldt 2010) in R.
The maximal number of tries for the delayed rejection was fixed to 1, so actually we used the adaptive Metropolis algorithm (Haario et al. 2001)\nocite{haario01}. The initial values were set by the \verb|modFit| function which finds the best fit parameters using optimization approaches. The variance of the proposal distribution was set by sample covariance of parameters $(x_1, \theta)$ scaled with $2.4^4/(p+q)$ in every 100 iteration. We got $10,000$ posterior sample from the DRAM algorithm. The DRAM was used here as a benchmark method for obtaining exact results based on Markov Chain Monte Carlo procedures but at the expense of computational time.

Gaussian processes (GP) have been used to avoid the heavy computation of the numerical integration. 
AGM (Dondelinger et al. 2013) and GP-ODE (Wang and Barber 2014) are two state-of-art paradigms for modelling the differential equation models using GP. 
The gradient matching (GM) approach (Calderhead et al. 2009)\nocite{calderhead2009accelerating} was developed to infer the differential equation models based on GP. 
However, GM approach has disadvantages that the posterior of hyperparameter of GP does not depend on the differential equation system and it is not a generative model. 
Dondelinger et al. (2013) tried to remedy the former problem by substituting the directed edges between the hyperparameter and the GP with the undirected edges. This modification improved the performance of the inference, but it is still not a generative model. 
GP-ODE approach was developed by Wang and Barber (2014) to construct a simple generative model. They developed a different paradigm from gradient matching approaches and argued the GP-ODE approach performs at least as well as the AGM.  
However, GP-ODE has been shown to be conceptually problematic. Recently, Macdonald et al. (2015)\nocite{macdonald2015controversy} pointed out that GP-ODE approach makes an undesirable approximation: GP-ODE eliminates the edge between the true state variable $x(t)$ and the latent variable $\tilde{x}(t)$ which should be same to the true state variable. 
Macdonald et al. (2015) showed that AGM achieves better result than GP-ODE for the simple ODE model having missing values and comparable result for the FitzHugh-Nagumo system.

To compare our LAP inference with the GP based approaches, we illustrated the results from both GP-ODE and AGM approaches. The MATLAB code for GP-ODE is available from github, and Macdonald et al. (2015)\nocite{macdonald2015controversy} provided the MATLAB code for the AGM approach. 
All parameters were sampled from griddy Gibbs sampling. The range for each parameter component $\theta_i$ was chosen by $[\widehat{\theta}_i^R \pm 4 \widehat{sd}(\widehat{\theta}_i^R)]$ where $\widehat{\theta}_i^R$ is the estimate from the parameter cascading method (Ramsay et al. 2007). We devided it into $31$ intervals of equal length to set the same number of grid for each parameter. For the variance function of GP, we chose squared exponential function $c_{\phi_j}(t,t') = \sigma_j^x \exp(- l_j(t-t')^2)$ and discretized the parameters $\sigma_j^x, l_j$ over the ranges $[0.1, 1], [5,50]$ with intervals $0.1, 5$, respectively.
We got $10,000$ posterior sample from the GP-ODE and AGM approaches. 

As we have concluded in the above simulation, we used the 4th order Runge-Kutta method with $m=2$ for the LAP inference and got $10,000$ posterior sample from each simulation data set.
The same grid set as GP-ODE was chosen for fair comparison.

\begin{table}[!tb]
	\scriptsize\centering\vspace{0cm}
	\caption{The table of mean of the absolute biases, the standard deviations (sd), the root mean squared errors (rmse) for $\hat{\theta}$, log-likelihoods with estimated parameters and computations times (s) in the FitzHugh-Nagumo model. The results for the Laplace approximated posterior (LAP) inference, parameter cascading (PC) method, delayed rejection adaptive Metropolis (DRAM) algorithm, GP-ODE approach and adaptive gradient matching (AGM) approach are shown. PC method with forward prediction error (FPE) criterion for the choice of $\lambda$ is denoted by PC FPE.}
	\hspace{.5cm}
	\begin{tabular}{cc|c|c|c|}\cline{3-5}
		\multicolumn{2}{c|}{} & LAP & PC & PC FPE   \\ \hline 
		\multicolumn{1}{|c}{\multirow{3}{*}{Absolute bias}} & \multicolumn{1}{|c|}{$\theta_1$} & 0.179   & 0.256   & 0.264    \\ 
		\multicolumn{1}{|c}{} & \multicolumn{1}{|c|}{$\theta_2$} & 0.222   & 0.246  & 0.257    \\
		\multicolumn{1}{|c}{} & \multicolumn{1}{|c|}{$\theta_3$} & 0.598 &  0.815 & 0.762 \\ \hline
		\multicolumn{1}{|c}{\multirow{3}{*}{sd}} & \multicolumn{1}{|c|}{$\theta_1$} & 0.220   & 0.290    & 0.369    \\ 
		\multicolumn{1}{|c}{} & \multicolumn{1}{|c|}{$\theta_2$} & 0.308   & 0.370    & 0.470    \\ 
		\multicolumn{1}{|c}{} & \multicolumn{1}{|c|}{$\theta_3$} & 0.679 & 0.825 & 0.913  \\ \hline
		\multicolumn{1}{|c}{\multirow{3}{*}{rmse}} & \multicolumn{1}{|c|}{$\theta_1$} & 0.298   & 0.493   & 0.488    \\
		\multicolumn{1}{|c}{} & \multicolumn{1}{|c|}{$\theta_2$} & 0.400   &  0.578    & 0.576    \\ 
		\multicolumn{1}{|c}{} & \multicolumn{1}{|c|}{$\theta_3$} & 0.954 & 1.299 & 1.290  \\ \hline
		\multicolumn{2}{|c|}{Log-likelihood}  & -7.128 & -7.059 & -7.665  \\ \hline
		\multicolumn{2}{|c|}{Computation time}  & 64.033 & 3.476 & 34.452  \\ \hline 
		\multicolumn{2}{|c|}{Software}  & R and Fortran90 & R and C/C++ & R and C/C++  \\ \hline \\ \cline{3-5}
		\multicolumn{2}{c|}{}  & DRAM & GP-ODE & AGM  \\ \hline 
		\multicolumn{1}{|c}{\multirow{3}{*}{Absolute bias}} & \multicolumn{1}{|c|}{$\theta_1$} & 0.239   & 0.159  & 0.457      \\ 
		\multicolumn{1}{|c}{} & \multicolumn{1}{|c|}{$\theta_2$} & 0.512   & 0.193  & 0.168   \\
		\multicolumn{1}{|c}{} & \multicolumn{1}{|c|}{$\theta_3$} & 0.654 & 1.439 & 1.842 \\ \hline
		\multicolumn{1}{|c}{\multirow{3}{*}{sd}} & \multicolumn{1}{|c|}{$\theta_1$} & 0.295   & 0.336  & 0.089   \\ 
		\multicolumn{1}{|c}{} & \multicolumn{1}{|c|}{$\theta_2$} & 0.621   & 0.411  & 0.267   \\ 
		\multicolumn{1}{|c}{} & \multicolumn{1}{|c|}{$\theta_3$} & 0.756 & 0.511 & 0.074 \\ \hline
		\multicolumn{1}{|c}{\multirow{3}{*}{rmse}} & \multicolumn{1}{|c|}{$\theta_1$} & 0.397   & 0.405  & 0.472   \\
		\multicolumn{1}{|c}{} & \multicolumn{1}{|c|}{$\theta_2$} & 0.833   & 0.472  & 0.333   \\ 
		\multicolumn{1}{|c}{} & \multicolumn{1}{|c|}{$\theta_3$} & 1.071 & 1.563 & 1.844 \\ \hline
		\multicolumn{2}{|c|}{Log-likelihood}  & -8.551 & -28.247 & -25.567 \\ \hline
		\multicolumn{2}{|c|}{Computation time}  & 85.327 & 6222.268 & 5235.615 \\ \hline
		\multicolumn{2}{|c|}{Software}  & R and C/C++ & MATLAB & MATLAB  \\ \hline
	\end{tabular}\label{table:GScompare} 
\end{table}
Table \ref{table:GScompare} shows the table of mean of the absolute biases, the standard deviations, the root mean squared errors (rmse) for $\hat{\theta}$, log-likelihoods with estimated parameters and computations times. The absolute bias term is calculated by
$$| \text{Bias}^s(\theta_i)| =  |\theta_i - \widehat{\theta}^s_i | , ~~ i = 1, 2,3$$
where $\text{Bias}^s$ is the bias in $s$-th simulation and $\widehat{\theta}^s$ is the estimate of $\theta$ in $s$-th simulation and $\theta = (0.2, 0.2, 3)^T$. For the Bayesian procedures, we use the posterior mean as the estimate of the parameter.

Table \ref{table:GScompare} shows that the LAP inference has better performance than the other methods in terms of rmse. 
The LAP has lower rmse and higher log-likelihood than those of the DRAM method, while taking $25\% $ less computational time compared to DRAM. 
The PC method with $\lambda=10^5$ has the fastest computation time and has a slightly higher log-likelihood value than that of LAP, while the automatic choice of $\lambda$ (PC FPE) has a comparable rmse and a relatively lower log-likelihood value as shown in Table \ref{table:GScompare}. 
The GP-ODE and AGM do not perform well in terms of the computational speed and accuracy (as determined by rmse).  

We have also checked the values of log-likelihood at the parameter estimates for each method which are shown in Table \ref{table:GScompare}.  Note that LAP consistently achieves the higher log-likelihood value corresponding to its parameter estimates (which is comparable to PC and DRAM) than GP-ODE and AGM.\ech Note that if there was a parameter estimate from another method, different from the LAP estimate but comparable in explaining the data, the log-likelihood at that parameter value (for the other method) should be close to the log-likelihood value corresponding to the LAP estimate. But Table \ref{table:GScompare} shows that in terms of the log-likelihood values, this is not the case;  GP-ODE and AGM yield significantly lower log-likelihood values suggesting suboptimal parameter estimates from them.  

To understand suboptimal parameter estimates, we note that the Fitz-Hugh-Nagumo ODE model has large regions of the likelihood corresponding to unidentifiable parameter values. However, this (large regions of the likelihood where parameter values are unidentifiable) does not arise
 for parameter values close to the maximum likelihood point (MLE) and for large sample size $n$. Our method based on Laplace approximation finds this maximum likelihood estimate and hence the bias is of order $O(n^{-1/2})$.  As seen from the log-likelihood values in Table \ref{table:GScompare}, GP-ODE and AGM give parameter estimates away from the MLE, and hence may belong to these regions of unidentifiability. In other words, parameter estimates from GP-ODE and AGM are genuinely deviating away from the true parameter value. Over repeated simulation experiments, these genuine deviations get translated into the large overall biases and standard deviations given in Table \ref{table:GScompare} (especially component $\theta_3$).

The coverage probabilities of 95\% credible interval of the LAP inference are comparable to those of the confidence intervals obtained by the other methods. The coverage probabilities for $\theta_1,\theta_2, \theta_3$ of the LAP inference are $0.94, 0.96, 0.94$, while those of the PC method and DRAM algorithm are $0.84, 0.91, 0.81$ and $0.96, 0.93, 0.97$, respectively.


\subsection{Predator-prey system}\label{PP}
Fussmann et al. (2000) suggested a mathematical model for predator-prey food chain between two microbials. \nocite{fussmann00}
 The following system  of equations describes the predator-prey oscillation between Brachionus calyciflorus and Chlorella vulgaris:
\bea
\dot{x}_1(t) &=& \delta ( N^* - x_1(t) ) - \frac{\theta_1 x_1(t) x_2(t)}{\theta_2 + x_1(t)}\\
\dot{x}_2(t) &=& \frac{\theta_1 x_1(t) x_2(t)}{\theta_2 + x_1(t)} - \frac{\theta_3 x_2(t)x_4(t)}{\theta_4 + x_2(t)}\cdot \frac{1}{\theta_5} - \delta x_2(t) \\
\dot{x}_3(t) &=& \frac{\theta_3 x_2(t) x_3(t)}{\theta_4 + x_2(t)} - ( \delta + \theta_6 + \theta_7) x_3(t) \\
\dot{x}_4(t) &=& \frac{\theta_3 x_2(t) x_3(t)}{\theta_4 + x_2(t)} - (\delta + \theta_6) x_4(t).
\eea
In the above model, $x_1, x_2, x_3, x_4$ represent the concentrations of nitrogen, Chlorella, reproducing Brachionus and total Brachionus, respectively. The unit of Chlorella and Brachionus is $\mu mol L^{-1}$.  $N^*$ is the inflow concentration of nitrogen, and $\delta$ is dilution rate.
We have seven positive parameters, $\theta= (\theta_1, \ldots, \theta_7)^T$.
$\theta_1$ and $\theta_2$ are the maximum birth rate and the half-saturation constant of Chlorella.
$\theta_3$ and $\theta_4$ represent the maximum birth rate and the half-saturation constant of Brachionus.
$\theta_5, \theta_6$, and $\theta_7$ are the assimilation efficiency, the mortality and the decay of fecundity of Brachionus.

The dimension of the parameter is 7 which is too big for  the grid sampling. Instead, we applied the griddy Gibbs sampling method.
We generated a simulated data set with model parameters $\theta = (3.3, 0.43, 2.25, 1.5, 2.5, 0.055, 0.4)^T$, $x_1 = (1, 3, 5, 5)^T$, $\sigma^2 = 1$ and $N^* = 8, \delta = 0.68$. We used the absolute value of the data because the concentrations should be positive.
The parameter settings come from Cao et al. (2008)\nocite{cao2008estimating}. We just modified the scale of $x_1, x_2, \theta_2, \theta_4, \theta_5$ and $N^*$ to control the scale of $x_1$ and $x_2$.
The time interval was fixed at $t_i - t_{i-1} = 0.1$ for $i = 2, 3, \ldots, n$ where $n=100$. We applied the 4th order Runge-Kutta method to get the true mean function for simulated data with $m=1$.

For the prior, we had $x_1 \mid \tau^2 \sim N_4(\mu_{x_1} = y_1, 100/\tau^2 I_4),$ $\tau^2 \sim Gamma(a,b)$ and $\theta \sim Unif(A)$ where $a=0.1, b=0.01$, $y_1 = (0.103, 3.185,  6.298,  5.137)^T$ and $A = \{ (\theta_1,\ldots, \theta_7) : 0< \theta_1, \theta_3, \theta_4, \theta_5 < 70, 0< \theta_2, \theta_6, \theta_7 < 10 \}$.

For this example, we ended up setting $M=15$ and $h_0 = (0.35, 0.40, 0.15, 0.17, 0.40, 0.07, 0.06)^T$, so we have $31$ grid points for each $\theta_j,~ j=1,\ldots, 7$.
The center of the grid matrix $\theta^0$ was chosen as $(3.295, 1.444, 2.225, 1.393, 3.883, 0.248,$ $0.397)^T$.

\begin{figure*}[!t]\centering	
		\includegraphics[width=.95\textwidth]{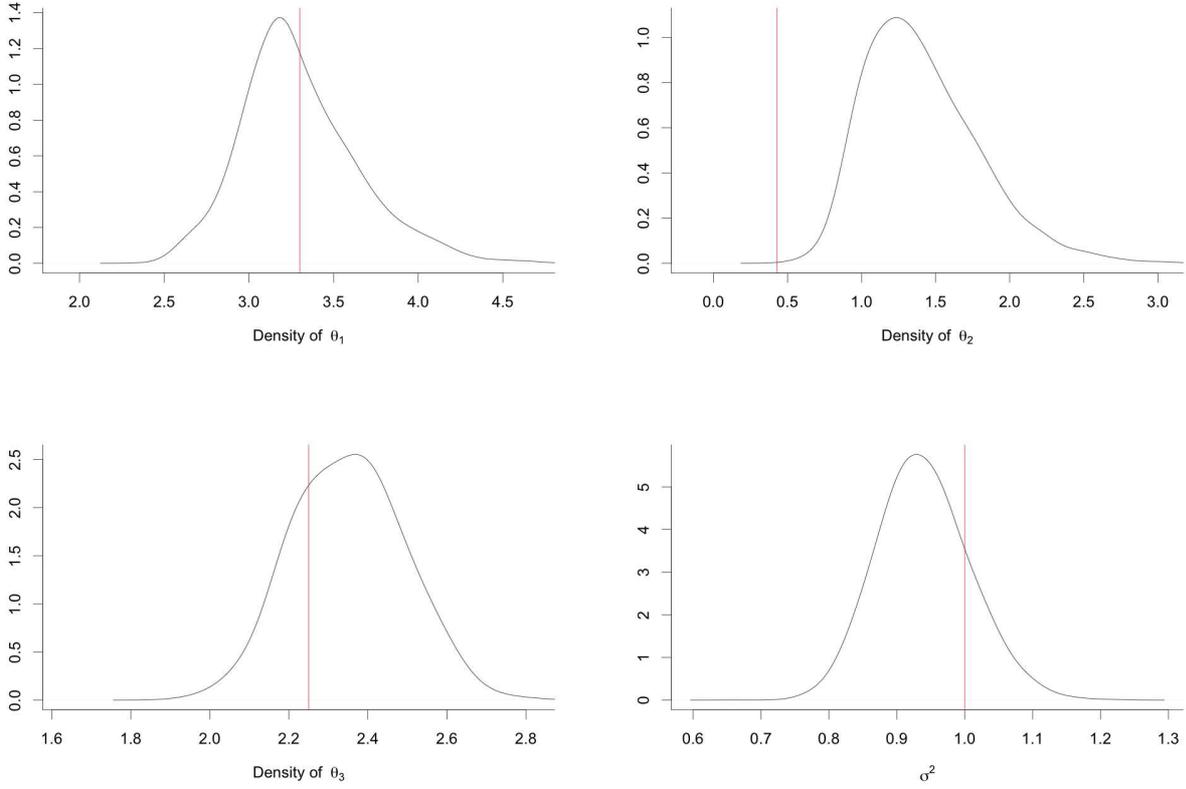}
		\caption{Approximate marginal posterior densities for $\theta_1$ (top left), $\theta_2$ (top right), $\theta_3$ (bottom left) and $\sigma^2$ (bottom right) for Predator-prey model. The step size $m=1$ in 4th order Runge-Kutta method is used. We omit the densities for the rest of the parameters, and the red lines represent true values of the parameters.}
	\label{fig:PPplot}			
\end{figure*}
\begin{table}[!b]
\caption{Posterior summary statistics with step size $m=1$ in 4th order Runge-Kutta method for Predator-prey model.}
\small\centering
\begin{tabular}{|l c c c c c|}
\hline
 & $\theta_1$ & $\theta_2$ & $\theta_3$ & $\theta_4$ & $\theta_5$ \\ \hline
True value & 3.3 & 0.43 & 2.25 & 1.5 &  2.5 \\
Mean & 3.298 &  1.410 &  2.351 & 1.214 & 3.634 \\
Median & 3.202 & 1.338 & 2.345 & 1.212 & 3.563 \\
5$\%$ quantile & 2.828 & 0.911 & 2.145 & 0.985 &  3.030 \\
95$\%$ quantile & 3.948 & 2.084 & 2.585 & 1.484 & 4.523  \\ \hline
 & $\theta_6$ & $\theta_7$ & $\sigma^2$ & & \\  \hline
True value  & 0.055 & 0.4  & 1 & & \\
Mean & 0.229 & 0.450 & 0.940 & & \\
Median & 0.211 & 0.445 & 0.937 & & \\
5$\%$ quantile  & 0.117 & 0.381 & 0.834 & &\\
95$\%$ quantile & 0.416 & 0.525 & 1.056 & & \\ \hline
\end{tabular}
\label{table:PPtb}
\end{table}

Total 50,000 posterior sample was drawn by the griddy Gibbs sampling and  every 5-th draw was used  as sample for the posterior inference; finally we got 10,000 posterior sample. It took 19.454 hours for this simulation.
Figure \ref{fig:PPplot} shows the approximate marginal posterior densities for some parameters. The summary statistics for the posterior is given at Table \ref{table:PPtb} with true value of the parameters $\theta$ and $\sigma^2$.

\begin{figure*}[!t]
		\centering
		\includegraphics[width=1.\textwidth]{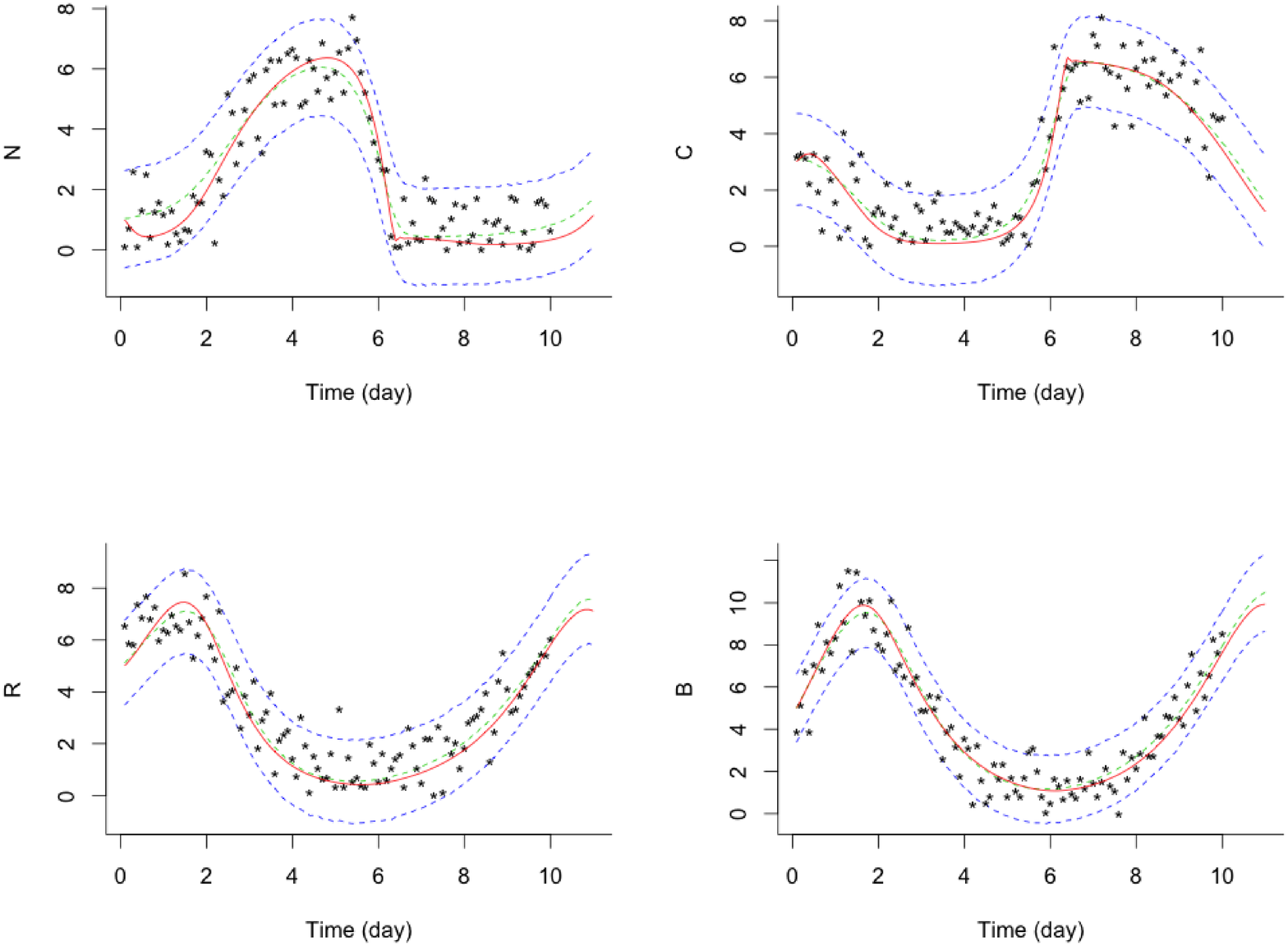}
		\caption{Scatter plot of the observations generated from the Predator-prey model, and plots of $90\%$ credible set lines and true $x(t)$ values are drawn when $m=1$. Predictions of 10 time points ahead are also drawn. The upper, lower and middle dotted lines are the 95\% and 5\% quantiles and mean of the posterior, respectively. The solid line in the middle is the true value of the state $x(t)$, and the star-shaped points are the observations.}
		\label{fig:PPpre}
\end{figure*}
Figure \ref{fig:PPpre} contains the scatter plots of the observations for $x_1, x_2, x_3, x_4$, the true mean functions, 90$\%$ credible lines for the mean functions, and the posterior mean functions as well as prediction values at 10 future time points.

As commented by one of the referees, we have adjusted the amount of error in our simulations to make the SNR (signal-to-noise-ratio) close to 10 to resemble real life situations. For the predator-prey model, the scale of some parameters were chosen to control the variance of signal. As a result, on the Newton's law of cooling model, the SNR for the different dataset sizes were obtained as follows: when $n=20,$ SNR $=10.493$, when $n=50,$ SRN $=8.313$, when $n=100,$ SNR $=7.660$, when $n=150,$ SNR $=7.450$. For the FitzHugh-Nagumo system, the SNR on species 1 is 8.598 and on species 2 is 1.928. For the predator-prey system, the SNRs on $x_1,x_2,x_3,x_4$ are $5.712, 6.112, 5.696, 8.369$, respectively.

\section{U.S. Census Data: logistic equation}\label{sec:6}
A simple logistic equation describing  the evolution of an animal population over time is
\begin{eqnarray}\label{LG}
{\dot x}(t) = \frac{\theta_1}{\theta_2} x(t) ({\theta_2} - x(t)),
\end{eqnarray}
where $x(t)$ is the population size at time $t$, $\theta_1$ is the rate of maximum population growth, $\theta_2$ is the maximum sustainable population sometimes called carrying capacity (Baca{\"e}r 2011)\nocite{bacaer2011verhulst}.  
The analytic form of the solution to (\ref{LG}) can be found. See Law et al. (2003) for the details.\nocite{Law03} In this example, however, we will use only the differential equation (\ref{LG}) to fit the model.

U.S. takes a census of its population every 10 years which is  mandated by the U. S. Constitution.
It has important ramifications for many aspects. For instance, the  census results are used in the decision of  government program funding, congressional  seat, and electoral votes.
This data set represents U.S. population from 1790 to 2010.
The population is represented by one million units.

Since the census has been conducted every 10 years from 1790 to 2010, we have total $n=23$ observations, $(y_1,\ldots, y_{23})$, with $h=t_i - t_{i-1} = 10$, $i=1,2,\ldots,n$.

We set the prior as $x_1 \mid \tau^2 \sim N(\mu_{x_1}=y_1, 100/\tau^2)$, $\tau^2 \sim Gamma(a, b)$ and $\theta \sim Uniform(0,1) \times Uniform(300, 1000)$, where $a=0.1, b=0.01$ and $y_1 = 3.929$.
The lower limit of $\theta_2$ was set to $300$ which is slightly lower than the population in year 2010, $y_{23} =308.746$.

To apply the grid sampling method, we used the parameter cascading estimate as a center of an initial grid set. For the final analysis, we set $M=35$, $h_0=(0.12, 0.4)^T$ and $\theta^0 = (0.020, 532.125)^T$.

\begin{figure*}[!t]
\begin{tabular}{ccc}
  \includegraphics[width=0.35\textwidth]{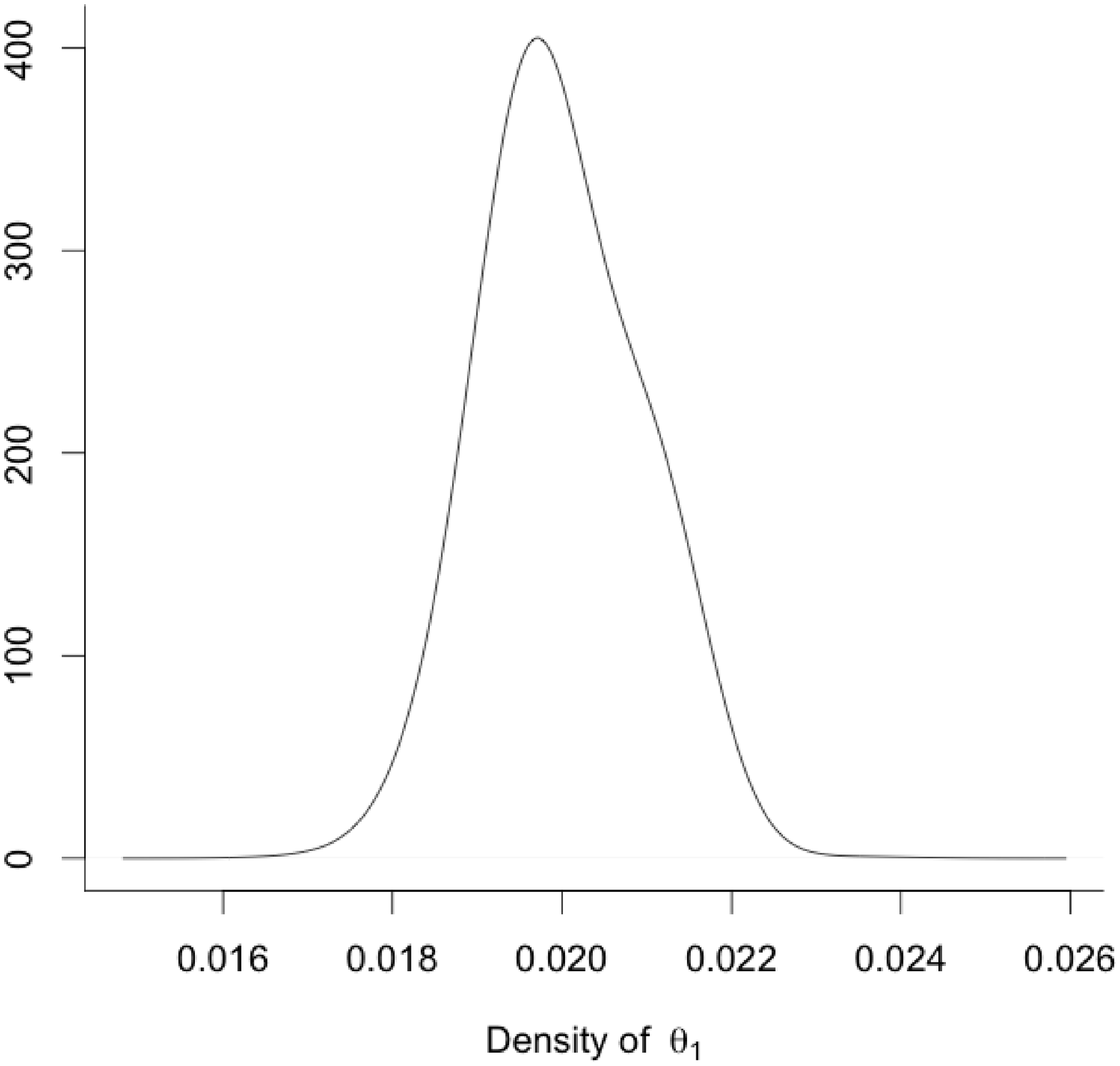} &\hspace{-1cm}
  \includegraphics[width=0.35\textwidth]{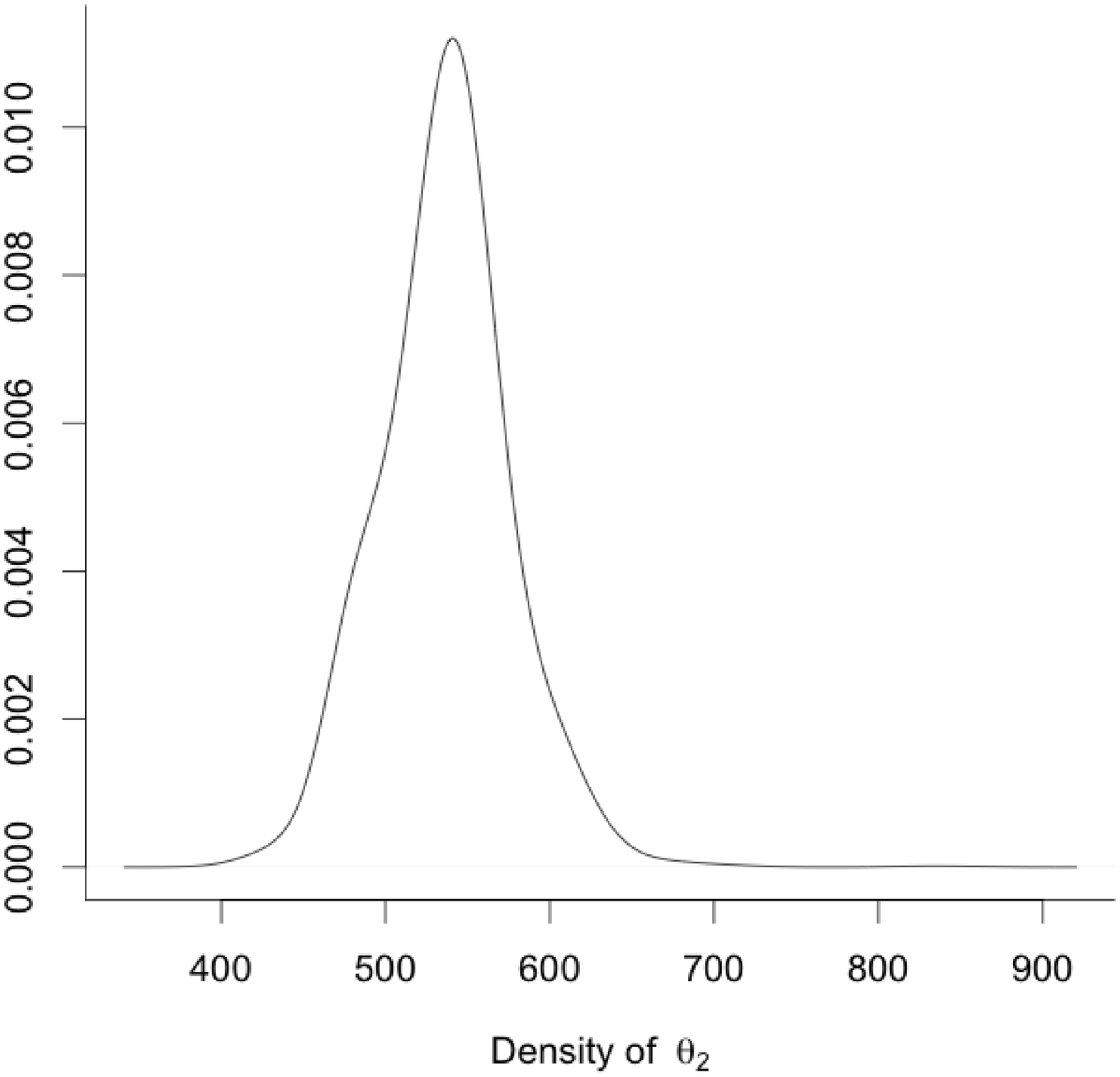} &\hspace{-1cm}
  \includegraphics[width=0.35\textwidth]{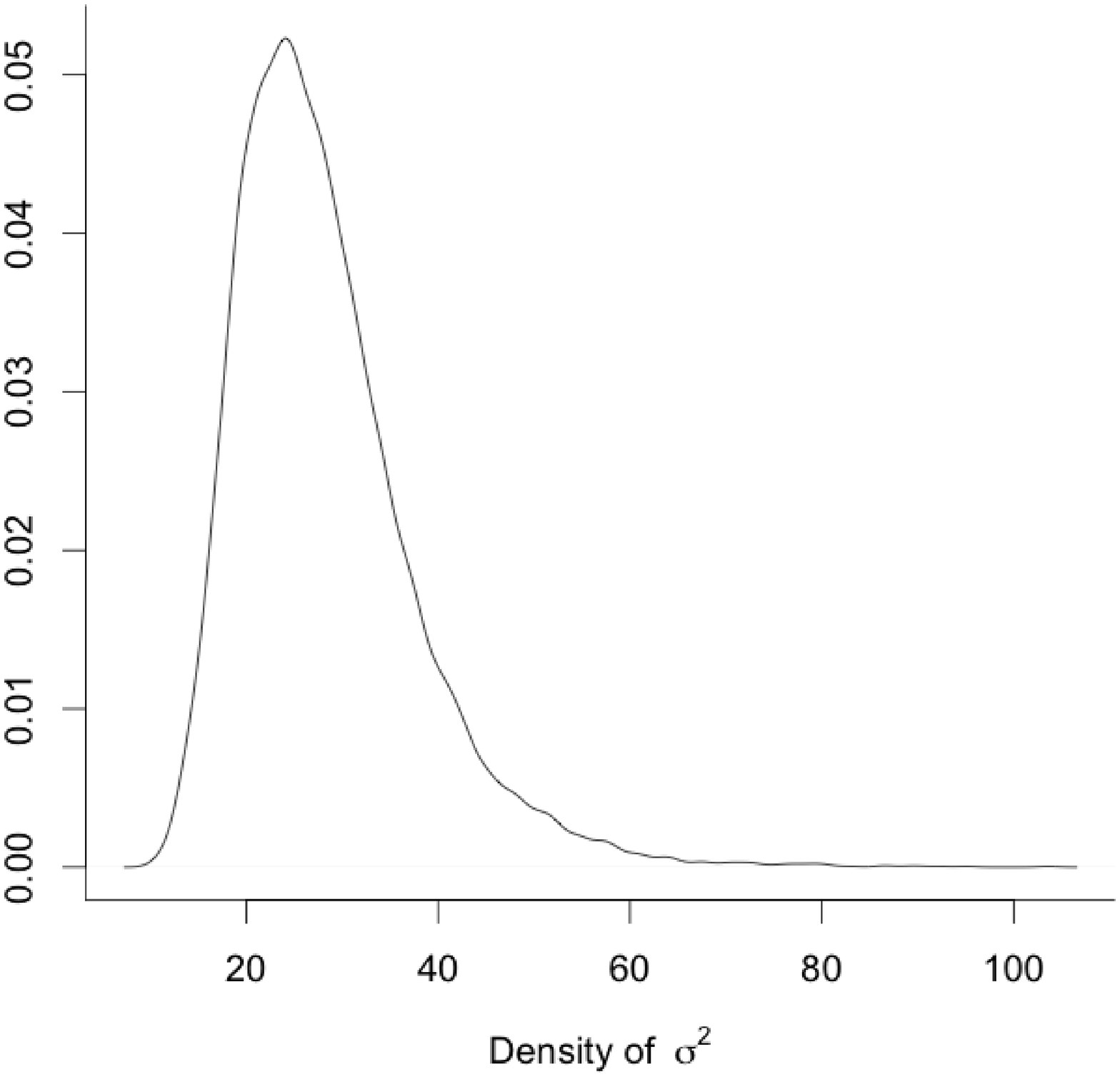}
\end{tabular}\vspace{-0.5cm}
\caption{The marginal posterior densities of $\theta_1$, $\theta_2$ and $\sigma^2$ in the logistic model with U.S. census data when the step parameter $m=1$.}\label{fig:LG}
\end{figure*}

\begin{figure*}[!htpb]	
\vspace{-0cm}
	\centering
		\includegraphics[width=.8\textwidth]{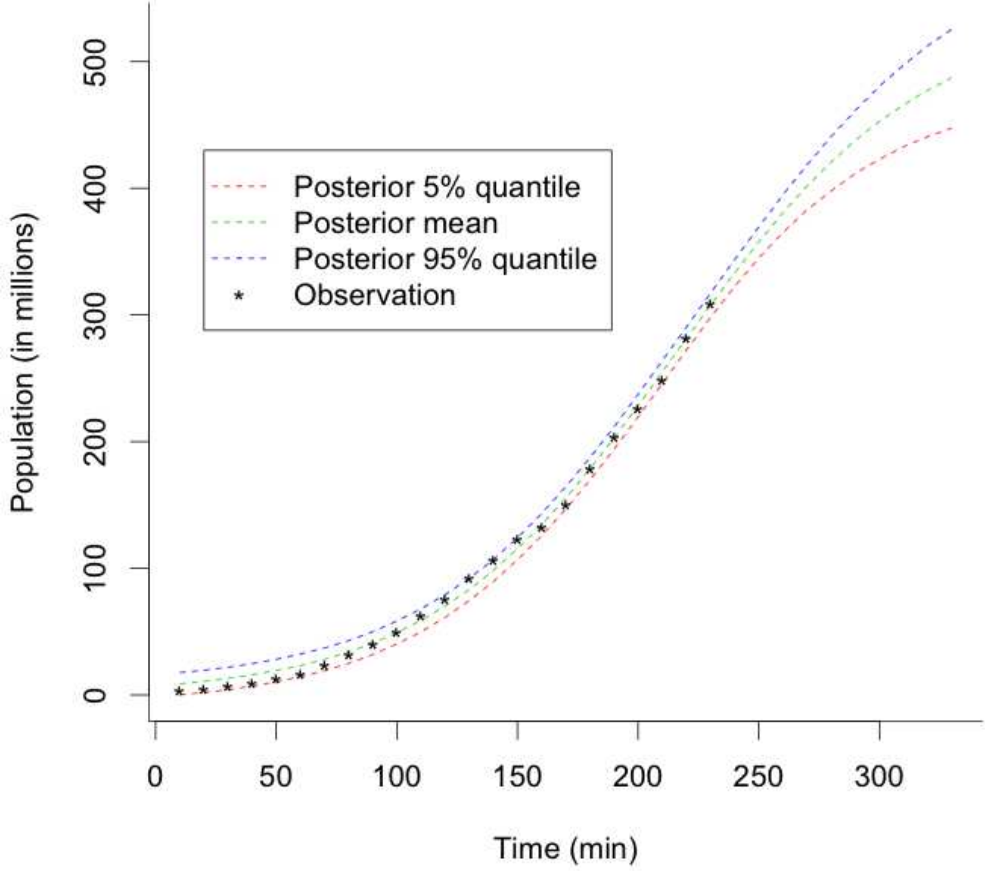}
		\vspace{-0.5cm}\caption{Scatter plot of the U.S. census, $90\%$ credible interval lines and posterior mean  when the step parameter $m=1$. Prediction values of populations at $10$ future time points are also drawn.}
	\label{fig:LGpred}
\end{figure*}

We tried several step sizes $m$ and concluded that with $m=1$ the posterior had been stabilized sufficiently. For the numerical approximation, we used the 4th order Runge-Kutta method. The marginal posterior densities of $\theta_1$, $\theta_2$ and $\sigma^2$ when $m=1$ are given in Figure \ref{fig:LG}.
Figure \ref{fig:LGpred} includes the scatter plot of the observations, the $90\%$ credible interval lines and posterior mean as well as prediction values of populations at $10$ future time points. Table \ref{table:LG} shows the summary statistics for the posterior.

\begin{table}[!tb]
\caption{Posterior summaries with $m=1$ for U.S. census data. C.I. denotes the credible interval.}
\scriptsize
\begin{tabular}{|c|c|c|c|}
  \hline
 &$\theta_1$ & $\theta_2$ & $\sigma^2$ \\ \hline
Mean   & 0.020 & 534.528  &  28.276 \\
Median & 0.020  & 532.125  & 26.314 \\
90\% C.I. &  (0.019, 0.021) & (482.817, 597.867) & (16.430, 46.367) \\ \hline
\end{tabular}
\label{table:LG}
\end{table}

\section{Discussion}\label{section:discussion}
In this paper, we proposed a posterior computation method, the LAP, based on the Laplace method and numerical approximation for ODE. There are three advantages of the proposed method.
First, when the dimension of the parameter is small, the computation is fast. The main issue of the proposed method is computation time when the dimension of $\theta$ is high.

Second, the proposed method produces accurate estimator which has comparable or better performance than the other methods:  the PC method, the DRAM, the GP-ODE and the AGM. Although it is not entirely clear, we suspect that the spline approximation of the PC method and the GP approximation of the GP based approaches to $x(t)$ may cause loss of efficiency. This issue also needs further investigation. 

Third, the proposed method is numerically stable. The frequentist methods need to maximize the log-likelihood surface which has many ripples. However, in many examples the ripples of the log-likelihood surface occurs at periphery of the parameter space and disappear in the likelihood surface as the sample size $n$ increases. 

Referees pointed that there is a potential to use lattice rule or sparse grid construction which can control the computational costs of the proposed method. 
It is an attractive way to reduce the computation time of LAP when $q$ is large. 
However, there were several challenges that need to be overcome. 
For the lattice rule, the best way of transforming integration domain to optimize its performance in the case of ODE models is not clear. It should be chosen carefully because poor transformation will cause the evaluations of ratios of densities under the lattice rule to be quite unstable. 
For the sparse grid, the existence of negative weights prevents computing the posterior probability on each grid point: we can compute the posterior moments only. To get the posterior probability on each grid point, the weights should be positive everywhere. Furthermore, in our experiment, the estimate from the sparse grid heavily depended on the range of the grid set and the accuracy of the sparse grid.
We applied the sparse grid construction to the proposed method for predator-prey system. The Gauss-Legendre quadrature rule on $[0,1]$ was used with accuracy level 10. The integration domain was transformed to the same domain in Sect. \ref{PP} using linear transformation.
In this settings, we obtained the mean value $(2.868, 1.281, 1.969, 1.225, 2.696, 0.116, 0.362)^T$ for $\theta$. The estimated mean or other moments were quite unstable to the choice of the domain and accuracy level.

Although we concluded that these problems are not easy to get around, the lattice rule and sparse grid are interesting idea to enhance the practical use of our LAP inference. 
Thus, we decided that applying the lattice rule or sparse grid to LAP inference deserves a separate investigation and publication.

\appendix

\section{Appendix}
\subsection{Computation of $\ddot{g}_n(x_1)$.}
Recall that
$$g_n(x_1) = \frac{1}{n} \sum_{i=1}^n \|y_i - x_i\|^2,$$
where $x_i = x(t_i)$ for $i=1,2,\ldots,n$ and $x(t) = (x_1(t), x_2(t),\ldots, x_p(t))^T$.
For the following discussion, we use the following convention for vectors and matrices. Suppose we have an array of real numbers $a_{ijk}$ with indices $i=1,2,\ldots, I$, $j=1,2,\ldots, J$ and $k=1,2,\ldots, K$.
Let $(a_{ijk})_{(i)}$ denote the column vector with dimension $I$
$$(a_{ijk})_{(i)} = (a_{1jk}, a_{2jk},\ldots, a_{Ijk})^T$$
and $(a_{ijk})_{(j,k)}$ denote the matrix with dimensions $J \times K$
$$(a_{ijk})_{(j,k)}= \left[\begin{array}{cccc}
			a_{i,1,1} &a_{i,1,2} & \ldots & a_{i,1,K} \\
			a_{i,2,1} & a_{i,2,2} & \ldots & a_{i,2,K} \\
			\ldots&\ldots&\ldots&\ldots \\
			a_{i,J,1} & a_{i,J,2} & \ldots & a_{i,J,K}
			\end{array} \right].
			$$
The indices in the the subscript with parenthesis are the indices running in the vector or the matrix. The object with one running index  is a column vector, while the object with two running indices a matrix where the first and the second running index are for the row and column, respectively.

Note that
$$g_n(x_1) = \frac{1}{n} \sum_{i=1}^n g_{n i}(x_1),$$
where $g_{ni}(x_1) = y_i^T y_i - 2 x_i^T y_i + x_i^T x_i$.
Thus,  the $(l,k)th$ element of $\ddot{g}_n(x_1)$ is
$$\frac{\partial^2 g_n}{\partial x_{1l} \partial x_{1k}} =  \frac{1}{n} \sum_{i=1}^n \frac{\partial^2 g_{ni}}{\partial x_{1l} \partial x_{1k}}.$$

Note
$$\frac{\partial g_{ni}}{\partial x_{1k}} = -2 \sum_{j=1}^p y_{ij} \frac{\partial x_{ij}}{\partial x_{1k}} + 2 \sum_{j=1}^p x_{ij} \frac{\partial x_{ij}}{\partial x_{1k}}$$
and
$$\frac{\partial^2 g_{ni}}{\partial x_{1l}\partial x_{1k}} =-2 \sum_{j=1}^p y_{ij} \frac{\partial^2 x_{ij}}{\partial x_{1l}\partial x_{1k}} + 2 \sum_{j=1}^p \left( \frac{\partial x_{ij}}{\partial x_{1l}} \frac{\partial x_{ij}}{\partial x_{1k}} + x_{ij} \frac{\partial^2 x_{ij}}{\partial x_{1l} \partial x_{1k}} \right).$$
The above equation can be written in a matrix form
\begin{eqnarray*}
(\frac{\partial^2 g_{ni}}{\partial x_{1l} \partial x_{1k}})_{(l,k)} &=&
 -2 \sum_{j=1}^p (\frac{\partial^2 x_{ij}}{\partial x_{1l} \partial x_{1k}})_{(l,k)} y_{ij}\\
&& +2 (\frac{\partial x_{ij}}{\partial x_{1l}})_{(l,j)} (\frac{\partial x_{ij}}{\partial x_{1k}})_{(j,k)}\\
&& +2 \sum_{j=1}^p (\frac{\partial^2 x_{ij}}{\partial x_{1l} \partial x_{1k}})_{(l,k)} x_{ij} \\
 & = & 2 (\frac{\partial x_{ij}}{\partial x_{1l}})_{(l,j)} (\frac{\partial x_{ij}}{\partial x_{1k}})_{(j,k)}\\
&& +2 \sum_{j=1}^p (\frac{\partial^2 x_{ij}}{\partial x_{1l} \partial x_{1k}})_{(l,k)} (x_{ij}-y_{ij}).
\end{eqnarray*}
Thus,
\begin{eqnarray*}
\ddot{g}_n(x_1) &=& \frac{2}{n} \sum_{i=1}^n \left((\frac{\partial x_{ij}}{\partial x_{1l}})_{(l,j)} (\frac{\partial x_{ij}}{\partial x_{1k}})_{(j,k)}  + \sum_{j=1}^p (\frac{\partial^2 x_{ij}}{\partial x_{1l} \partial x_{1k}})_{(l,k)} (x_{ij}-y_{ij})\right).
\end{eqnarray*}

The derivatives of $x_i$ with respect to $x_1$ can be computed by using the sensitivity equation for ODE. See Hooker (2009).\nocite{Hooker09}
 Let
\bea
z_{jl}(t) = \frac{\partial x_j(t)}{\partial x_{1l}} \text{ ~or~ } Z(t) = \left(\frac{\partial x_j(t)}{\partial x_{1l}} \right)_{(j, l)} , ~~ j,l =1,\ldots, p
\eea
be the sensitivity of the state $x_{j}$ with respect to the initial value $x_{1l}$. The sensitivity equation is given by
\bea
\dot{z}_{jl}(t) = \frac{\partial}{\partial t}\frac{\partial x_j(t)}{\partial x_{1l}} &=& \frac{\partial}{\partial x_{1l}} \dot{x}_j(t)\\
&=& \sum_{u=1}^p \frac{\partial f_j( x, t ; \theta)}{\partial x_u(t)} \frac{\partial x_u(t)}{\partial x_{1l}}\\
&=& \sum_{u=1}^p \frac{\partial f_j( x, t ; \theta)}{\partial x_u(t)} z_{ul}(t),
\eea
or in matrix notation,
\begin{equation}\label{seneq}
\dot{Z}(t) = \left(\frac{\partial f_j( x, t ; \theta)}{\partial x_u(t)} \right)_{(j, u)} \cdot Z(t)
\end{equation}
with an initial condition $Z(t_1) = I_p$. For given $\theta$ and $t$, the coefficient $\partial f_j( x, t ;\theta) / \partial x_u(t)$ is calculated easily. It is a linear ODE problem whose initial condition is known. We can solve \eqref{seneq} using some numerical methods such as Runge-Kutta method.
$\partial^2 x_{ij} / (\partial x_{1l} \partial x_{1k})$ can be computed similarly.

\subsection{Proof of Theorem \ref{Conv}}
\begin{proof}
The results of Tierney and Kadane (1986) and Azevedo-Filho and Shachter (1994) assume several regularity conditions such as the existence of a unique global maximum as well as the existence of higher order derivatives (up to sixth order) of the likelihood function. In particular, our methods for approximating the ODE model work only under the assumption of a unique maximum of the likelihood function. Thus, we assume that the likelihood surface does not include any ridges (that is, continuum regions with equal values of the global maximum).
	
Using the result in Tierney and Kadane (1986) and Azevedo-Filho and Shachter (1994), we have
\bea
\pi_m(\theta, \tau^2 \mid {\bf y}_n) &=&  c_m^{-1} \int L_m(\theta, \tau^2, x_1) \pi(\theta,\tau^2,x_1) dx_1 \\
&& \times (1+ O(n^{-3/2})),
\eea
where  $c_m = \int L_m(\theta,\tau^2,x_1)\pi(\theta,\tau^2,x_1) dx_1 d\theta d\tau^2$.
Note the full likelihood $L(\theta, \tau^2, x_1)$ is
$$L(\theta, \tau^2, x_1) \propto e^{-\frac{\tau^2}{2}ng_n(x_1)} \times (\tau^2)^{np/2} ,$$
and $L_m(\theta, \tau^2, x_1)$ is the corresponding term with $g_n$ replaced by $g_n^m$.
If $L_m(\theta,\tau^2,x_1)$ converges to $L(\theta,\tau^2,x_1)$ as $m \to \infty$ for all $\theta \in \Theta, \tau^2 >0, x_1 \in \mathbb{R}^p$ and ${\bf y}_n$,
by the dominated convergence theorem, $c_m \longrightarrow c$ as $m \to \infty$. Thus,
\bea
\lim_{m\rightarrow\infty} \pi_m(\theta,\tau^2 \mid {\bf y}_n) & =& c^{-1} \int L(\theta,\tau^2,x_1)\pi(\theta,\tau^2,x_1)dx_1 \\
&& \times (1+ O(n^{-3/2}))\\
&=& \pi(\theta,\tau^2 \mid {\bf y}_n)\times (1+ O(n^{-3/2}))
\eea
which is the desired result.

To complete the proof,
we need to prove $L_m(\theta,\tau^2,x_1) \longrightarrow L(\theta,\tau^2,x_1)$ as $m\to \infty$, and it suffices to prove $ng_n^m(x_1) \longrightarrow ng_n(x_1)$ as $m\to\infty$.
Since we assume the Lipschitz continuity of $f$, the ODE has a unique solution with initial condition $x(t_1) = x_1$. Assumptions A1 and A3 implies
$$ \sup_{x, t} \| \frac{d^K}{dt^K} f(x,t;\theta) \| =: B < \infty$$
for some constants $B >0$.
The  local errors of the $K$th order numerical method are given by
$$ \| x(t_i) - x(t_{i-1}) - h \phi(x_{i-1} , t_{i-1} ; \theta) \|  \le B' h^{K+1}, ~ i=2,\ldots, n$$
for some $B'>0$, which
depends only on $\sup_{ t} \| d^K f(x,t;\theta)/ (dt^K) \|$ $\le B$ (Palais and Palais, 2009).\nocite{palais2009differential}
Thus, the  local errors are uniformly bounded. It implies the global errors uniformly bounded
$$\|x_i - x^h_i\| \le C h^K $$
for some constant $C>0$.

Thus,
\begin{eqnarray}\label{ngnrate}
| ng_n(x_1) - ng_n^m(x_1)| &=& \big| \sum_{i=1}^n \|y_i - x_i \|^2 - \sum_{i=1}^n\|y_i - x_i^m\|^2 \big| \nonumber \\
& = & \sum_{i=1}^n \big(  \|y_i - x_i \| +  \|y_i - x_i^m \| \big) \nonumber\\
&& \times  \big|  \|y_i - x_i \| -  \|y_i - x_i^m \| \big| \nonumber \\
&\le& \sum_{i=1}^n \big( 2 \|y_i - x_i \| + \|x_i -x_i^m\| \big) \| x_i-x_i^m\| \nonumber \\
&\le& \sum_{i=1}^n \big( 2C_y + 2 C_x + \|x_i -x_i^m\| \big) \|x_i -x_i^m\| \nonumber \\
&\le&\sum_{i=1}^n \left(  2C_y + 2 C_x + C \Big(\frac{h}{m}\Big)^K \right) C \Big(\frac{h}{m}\Big)^K  \nonumber \\
&\asymp& n \Big(\frac{h}{m} \Big)^K, \text{ as } m \longrightarrow \infty,
\end{eqnarray}
where  $\sup_{t \in [T_0,T_1]}\|y(t)\| < C_y < \infty$, $\sup_{t \in [T_0,T_1]}\|x(t)\| < C_x <\infty$.
This completes the proof.
\end{proof}

\subsection{Proof of Theorem \ref{mrate}}
\begin{proof}
If $\alpha > 5/(2K)$, as $n$ goes to infinity, $n(h/m)^K = O(n^{1-\alpha K}) = O(n^{-3/2})$ and it converges to  to zero.
Under $A1-A3$, we have shown in the proof of Theorem \ref{Conv} that  $| ng_n(x_1) - ng_n^m(x_1)| = O(n (h/m)^K)$.
For fixed $\tau^2>0$,
\bea
e^{-\frac{\tau^2}{2}ng_n^m(x_1)} &=& e^{-\frac{\tau^2}{2}[ng_n(x_1)+ng_n^m(x_1)-ng_n(x_1)]}\\
&=&  e^{-\frac{\tau^2}{2}ng_n(x_1)} \times e^{-\frac{\tau^2}{2}[ng_n^m(x_1)-ng_n(x_1)]} \\
&=& e^{-\frac{\tau^2}{2}ng_n(x_1)} \times e^{-\frac{\tau^2}{2}O(n (\frac{h}{m})^K)} \\
&=& e^{-\frac{\tau^2}{2}ng_n(x_1)} \times \left( 1+ O\Big(n \Big(\frac{h}{m}\Big)^K\Big) \right)
\eea
because $e^{x} = 1+ O(x)$ for sufficiently small $x$. It implies
\begin{eqnarray}\label{prf2}
\pi_m(\theta, \tau^2 \mid {\bf y}_n) &\propto&  \int L_m(\theta, \tau^2, x_1) \pi(\theta,\tau^2,x_1) dx_1 \times (1+ O(n^{-3/2})) \nonumber \\
&=&  \int L(\theta, \tau^2, x_1) \pi(\theta,\tau^2,x_1) dx_1 \times (1+ O(n^{-3/2}))\nonumber \\
&& \times \left( 1+ O\Big(n \Big(\frac{h}{m}\Big)^K \Big) \right)~~~\nonumber \\
&\propto& \pi(\theta,\tau^2\mid {\bf y}_n) \times (1+ O(n^{-3/2})) \times \left( 1+ O\Big(n \Big(\frac{h}{m}\Big)^K \Big) \right) \nonumber
\end{eqnarray}
for sufficiently large $n$. If $\alpha > 5/(2K)$, i.e., $n(h/m)^K \le n^{-3/2}$, $ (1+ O(n^{-3/2})) \times ( 1+ O(n (h/m)^K) )$  is $ (1+ O(n^{-3/2}))$. 
\end{proof}

\bibliographystyle{spbasic}
\bibliography{LAP}

\end{document}